\documentclass[twocolumn]{aastex631}

\usepackage{amsmath}
\usepackage{color}

\received{8-Sep-2022}
\revised{21-Dec-2022}
\accepted{23-Jan-2023}

\submitjournal{ApJ}

\shorttitle{SN2020uem: (I) The Nature of Type IIn/Ia-CSM SNe}
\shortauthors{UNO, MAEDA, \& NAGAO et al.}

\graphicspath{{./}{figures/}}

\begin{document}

\title{SN 2020uem: A Possible Thermonuclear Explosion within A Dense Circumstellar Medium\\
(I) The Nature of Type IIn/Ia-CSM SNe from Photometry and Spectroscopy}

\correspondingauthor{Kohki Uno}
\email{k.uno@kusastro.kyoto-u.ac.jp}

\author[0000-0002-6765-8988]{Kohki Uno}
\affiliation{Department of Astronomy, Kyoto University, Kitashirakawa-Oiwake-cho, Sakyo-ku, Kyoto, 606-8502, Japan}

\author[0000-0003-2611-7269]{Keiichi Maeda}
\affiliation{Department of Astronomy, Kyoto University, Kitashirakawa-Oiwake-cho, Sakyo-ku, Kyoto, 606-8502, Japan}

\author[0000-0002-3933-7861]{Takashi Nagao}
\affiliation{Department of Physics and Astronomy, University of Turku, FI-20014 Turku, Finland}

\author{Tatsuya Nakaoka}
\affiliation{Hiroshima Astrophysical Science Center, Hiroshima University, Kagamiyama 1-3-1, Higashi-Hiroshima ,Hiroshima 739-8526, Japan}
\affiliation{Department of Physical Science, Hiroshima University, Kagamiyama 1-3-1, Higashi-Hiroshima 739-8526, Japan}

\author[0000-0002-0724-9146]{Kentaro Motohara} 
\affiliation{National Astronomical Observatory, 2-21-1 Osawa, Mitaka, Tokyo 181-8588, Japan}

\author[0000-0001-8813-9338]{Akito Tajitsu} 
\affiliation{Okayama Branch Office, Subaru Telescope, National Astronomical Observatory of Japan, Kamogata, Asakuchi, Okayama, 719-0232, Japan}

\author{Masahito Konishi} 
\affiliation{Institute of Astronomy, Graduate School of Science, The University of Tokyo, 2-21-1 Osawa, Mitaka,
Tokyo 181-0015, Japan}

\author{Shuhei Koyama} 
\affiliation{Institute of Astronomy, Graduate School of Science, The University of Tokyo, 2-21-1 Osawa, Mitaka,
Tokyo 181-0015, Japan}

\author{Hidenori Takahashi} 
\affiliation{Institute of Astronomy, Graduate School of Science, The University of Tokyo, 2-21-1 Osawa, Mitaka,
Tokyo 181-0015, Japan}
\affiliation{Kiso Observatory, Institute of Astronomy, Faculty of Science, the University of Tokyo, Mitake 10762-30, Kiso-machi, Kiso-gun, Nagano 397-0101, Japan}

\author[0000-0001-8253-6850]{Masaomi Tanaka}
\affiliation{Astronomical Institute, Tohoku University, Sendai 980-8578, Japan}
\affiliation{Division for the Establishment of Frontier Sciences, Organization for Advanced Studies, Tohoku University, Sendai 980-8577, Japan}

\author[0000-0002-1132-1366]{Hanindyo Kuncarayakti}
\affiliation{Department of Physics and Astronomy, University of Turku, FI-20014 Turku, Finland}
\affiliation{Finnish Centre for Astronomy with ESO (FINCA), FI-20014 University of Turku, Finland}

\author[0000-0002-4540-4928]{Miho Kawabata}
\affiliation{Okayama Observatory, Kyoto University, 3037-5 Honjo, Kamogatacho, Asakuchi, Okayama 719-0232, Japan}

\author[0000-0001-9456-3709]{Masayuki Yamanaka}
\affiliation{Okayama Observatory, Kyoto University, 3037-5 Honjo, Kamogatacho, Asakuchi, Okayama 719-0232, Japan}

\author[0000-0003-4569-1098]{Kentaro Aoki} 
\affiliation{Subaru Telescope, National Astronomical Observatory of Japan, 650 North A’ohoku Place, Hilo, HI 96720, USA}

\author{Keisuke Isogai}
\affiliation{Okayama Observatory, Kyoto University, 3037-5 Honjo, Kamogatacho, Asakuchi, Okayama 719-0232, Japan}
\affiliation{Department of Multi-Disciplinary Sciences, Graduate School of Arts and Sciences, The University of Tokyo, 3-8-1 Komaba, Meguro, Tokyo 153-8902, Japan}

\author[0000-0002-8482-8993]{Kenta Taguchi}
\affiliation{Department of Astronomy, Kyoto University, Kitashirakawa-Oiwake-cho, Sakyo-ku, Kyoto, 606-8502, Japan}

\author[0000-0001-5822-1672]{Mao Ogawa}
\affiliation{Department of Astronomy, Kyoto University, Kitashirakawa-Oiwake-cho, Sakyo-ku, Kyoto, 606-8502, Japan}

\author[0000-0001-6099-9539]{Koji S. Kawabata}
\affiliation{Hiroshima Astrophysical Science Center, Hiroshima University, Kagamiyama 1-3-1, Higashi-Hiroshima ,Hiroshima 739-8526, Japan}
\affiliation{Department of Physical Science, Hiroshima University, Kagamiyama 1-3-1, Higashi-Hiroshima 739-8526, Japan}

\author{Yuzuru Yoshii} 
\affiliation{Institute of Astronomy, Graduate School of Science, The University of Tokyo, 2-21-1 Osawa, Mitaka,
Tokyo 181-0015, Japan}
\affiliation{Steward Observatory, University of Arizona, 933 North Cherry Avenue, Rm. N204 Tucson, AZ 85721-0065, USA}

\author{Takashi Miyata} 
\affiliation{Institute of Astronomy, Graduate School of Science, The University of Tokyo, 2-21-1 Osawa, Mitaka,
Tokyo 181-0015, Japan}

\author[0000-0002-0643-7946]{Ryo Imazawa}
\affiliation{Department of Physics, Graduate School of Advanced Science and Engineering, Hiroshima University, Kagamiyama 1-3-1, Higashi-Hiroshima, Hiroshima 739-8526, Japan}

\begin{abstract}

We have performed intensive follow-up observations of a Type IIn/Ia-CSM SN (SN IIn/Ia-CSM), 2020uem, with photometry, spectroscopy, and polarimetry. In this paper, we report on the results of our observations focusing on optical/near-infrared (NIR) photometry and spectroscopy. The maximum V-band magnitude of SN 2020uem is over $-19.5$ mag. The light curves decline slowly with a rate of $\sim 0.75 {\rm ~mag}/100 {\rm ~days}$. In the late phase ($\gtrsim 300$ days), the light curves show accelerated decay ($\sim 1.2 {\rm ~mag}/100 {\rm ~days}$). The optical spectra show prominent hydrogen emission lines and broad features possibly associated with Fe-peak elements. In addition, the $\rm H\alpha$ profile exhibits a narrow P-Cygni profile with the absorption minimum of $\sim  100 {\rm ~km~s^{-1}}$. SN 2020uem shows a higher $\rm H\alpha/H\beta$ ratio ($\sim 7$) than those of SNe IIn, which suggests a denser CSM. The NIR spectrum shows the Paschen and Brackett series with continuum excess in the H and Ks bands. We conclude that the NIR excess emission originates from newly-formed carbon dust. The dust mass ($M_{\rm d}$) and temperature ($T_{\rm d}$) are derived to be $(M_{\rm d}, T_{\rm d}) \sim  (4-7 \times 10^{-5} {\rm ~M_{\odot}}, 1500-1600 {\rm ~K})$. We discuss the differences and similarities between the observational properties of SNe IIn/Ia-CSM and those of other SNe Ia and interacting SNe. In particular, spectral features around $\sim 4650$ {\text \AA} and $\sim 5900$ {\text \AA} of SNe IIn/Ia-CSM are more suppressed than those of SNe Ia; these lines are possibly contributed, at least partly, by \ion{Mg}{1}] and \ion{Na}{1}, and may be suppressed by high ionization behind the reverse shock caused by the massive CSM.

\end{abstract}


\section{Introduction} \label{sec:1}

Type Ia supernovae \citep[SNe Ia;][]{Filippenko1997ARAA} are one of the most luminous transient events in the Universe. For photometric properties of typical SNe Ia, the peak magnitude of $\lesssim -19 {\rm ~mag}$ is reached $\sim 20$ days after the explosion. After the peak, the light curves quickly decline with a duration of $\sim 30 $ days before the decay is slowed down, i.e., the typical timescale of SNe Ia is $\sim 50$ days. As for spectroscopic properties, the early-phase spectra of SNe Ia are characterized by the absence of hydrogen and helium lines, and the presence of broad absorption features of silicon and sulfur with $\sim 10000 {\rm ~km~s^{-1}}$. In the nebular phase ($\gtrsim 100$ days), the spectra exhibit forbidden lines of iron and cobalt. 

As a theoretical interpretation, SNe Ia are triggered by thermonuclear explosions of white dwarfs \citep[e.g.,][for a review]{Maeda2016IJMPD}, which are end products of low-mass stars ($\lesssim 8 {\rm ~M_{\odot}}$). The main energy source of the SN Ia light curves is the radioactive decay of synthesized $\rm ^{56}Ni$ to $\rm ^{56}Co$, then $\rm ^{56}Fe$; the timescale at the peak matches the diffusion in the ejecta with the mass of $\sim 1 {\rm ~M_{\odot}}$ and the late-time light curve follows the quasi-exponential decay expected in this scenario. Besides, the white-dwarf thermonuclear explosion scenario explains the characteristic spectra of SNe Ia as originating from the Fe-rich ejecta.

In general, SNe Ia show relatively homogeneous properties in their light curves and spectra \citep[e.g.,][]{Phillips1993ApJ, Taubenberger2017SNhandbook}, although some SNe Ia show peculiar observational properties \citep[e.g.,][]{Li2003PASP}. Thanks to new generation surveys such as ASAS-SN \citep[All-Sky Automated Survey for SuperNovae;][]{Shappee2014AAS}, ATLAS \citep[Asteroid Terrestrial-impact Last Alert System;][]{Tonry2018PASP} and ZTF \citep[Zwicky Transient Facility;][]{Kulkarni2018ATel}, the number of rare variants of SNe Ia that show different observational features from the classical SNe Ia has been rapidly increasing recently \citep[e.g.,][]{Foley2013ApJ,Kawabata2018PASJ,Howell2006Nature,Yamanaka2009ApJ,Yamanaka2016PASJ,Perets2010Nature,Jacobson-Galan2020ApJ,Nakaoka2021ApJ}. It is becoming clear that SNe Ia have a large diversity in their photometric and spectroscopic properties \citep{Taubenberger2017SNhandbook}. 

A rare subclass of SNe, which is referred to as Type Ia-CSM SNe (SNe Ia-CSM), has been discovered \citep[e.g.,][]{Hamuy2003Nature,Dilday2012Science}. SNe Ia-CSM are characterized by their spectra, which exhibit prominent narrow hydrogen and helium emission lines, on top of spectra characterized by intermediate-mass element lines (i.e., Mg, Si, and Fe) similar to typical SNe Ia. Besides, for photometric properties, SNe Ia-CSM keep high luminosity for a long period ($\gtrsim 1 {\rm ~yr}$). The narrow emission lines and high luminosity suggest presence of a massive hydrogen-rich circumstellar medium (CSM) similar to Type IIn SNe (SNe IIn). The channel leading to SNe Ia-CSM still remains unclear; some suggest that SNe Ia-CSM are triggered by a thermonuclear explosion of a white dwarf in a dense CSM \citep[e.g.,][]{Deng2004ApJ, Fox2015MNRAS}, while others suggest that they originate from a CSM-interacting core-collapse SN \citep[e.g.,][]{Benetti2006ApJ}. Furthermore, frequently consensus has not been reached for their classification either as SN Ia-CSM or IIn for the SN Ia-CSM `candidates', and such events are termed as SNe IIn/Ia-CSM; their progenitor origin is even less clear \citep[e.g.,][]{Inserra2014MNRAS,Inserra2016MNRAS}. The sample of well-observed SNe IIn/Ia-CSM (including the genuine SNe Ia-CSM like PTF11kx and SN 2002ic) is still limited; e.g., SN 2002ic \citep{Hamuy2003Nature, Deng2004ApJ, Wang2004ApJ}, SN 2005gj \citep{Aldering2006ApJ, Prieto2007arxiv}, SN 2008J \citep{Taddia2012AA}, PTF11kx \citep{Dilday2012Science}, SN 2012ca \citep{Inserra2014MNRAS,Fox2015MNRAS,Inserra2016MNRAS}, and SN 2013dn \citep{Fox2015MNRAS}. To clarify the nature and origin of SNe IIn/Ia-CSM, we need to obtain a large sample of SNe IIn/Ia-CSM and compare their observational properties with other interacting SNe, e.g., SNe IIn or superluminous SNe.

As an interesting avenue, NIR spectroscopy plays potentially a key role in exploring the nature of CSM properties of SNe IIn/Ia-CSM. Interacting SNe are suggested to be one of the major factories of cosmic dust grains. Thanks to the dust thermal emission, some interacting SNe show high luminosity in the NIR wavelengths in the late phases. Indeed, NIR-luminous SNe IIn produce a large amount of dust \citep[$\gtrsim 10^{-4} {\rm ~M_{\odot}}$, e.g.,][]{Stritzinger2012ApJ,Maeda2013ApJ,Gall2014Nature}. However, two possibilities remain for the origin of the dust grains: pre-existing dust or newly-formed dust. For the pre-existing dust, which is formed by mass ejection before the explosions, e.g., stellar winds, the NIR emission may be produced by echoes. On the other hand, newly-formed dust could be formed by a cold-dense shell created by radiative cooling within a shocked CSM, and they may reprocess the SN optical photons to the NIR thermal emission. Although we expect that constraining on the origin of dust improves our understanding of the CSM and the progenitor system of SNe IIn/Ia-CSM, the sample of NIR observations, especially spectroscopy, is very limited for SNe IIn/Ia-CSM \citep[e.g.,][]{Fox2015MNRAS,Inserra2016MNRAS}.

We have performed intensive follow-up observations of SN IIn/Ia-CSM 2020uem for about 400 days after the discovery. In addition to photometry and optical spectroscopy, we performed high-dispersion spectroscopy, NIR spectroscopy, imaging polarimetry, and spectropolarimetry. These observations form one of the largest datasets for an SN IIn/Ia-CSM.

This paper presents the results of our optical/NIR photometric and spectroscopic observations. The polarimetry data are presented in Paper II. This paper is structured as follows. In Section \ref{sec:2}, we introduce basic information on SN 2020uem, followed by summaries of our observations and reduction processes. In Section \ref{sec:3}, we show multi-band and quasi-bolometric light curves. We compare the bolometric light curve with other interacting SNe and SNe Ia. In Section \ref{sec:4}, we analyze the optical and NIR spectra. In Section \ref{sec:5}, we estimate the dust mass for SN 2020uem and discuss the origin of the dust. In Section \ref{sec:6}, we discuss differences and similarities between the observational properties of SNe IIn/Ia-CSM, especially SN 2020uem, and those of other interacting SNe and SNe Ia. Besides, we discuss implications for the spectral formation based on the spectral comparison. This paper is closed in Section \ref{sec:7} with conclusions.

\section{Observations and data reduction} \label{sec:2}

\subsection{discovery of SN 2020uem} \label{sec:2.1}

\begin{figure*}
\epsscale{1.17}
\plotone{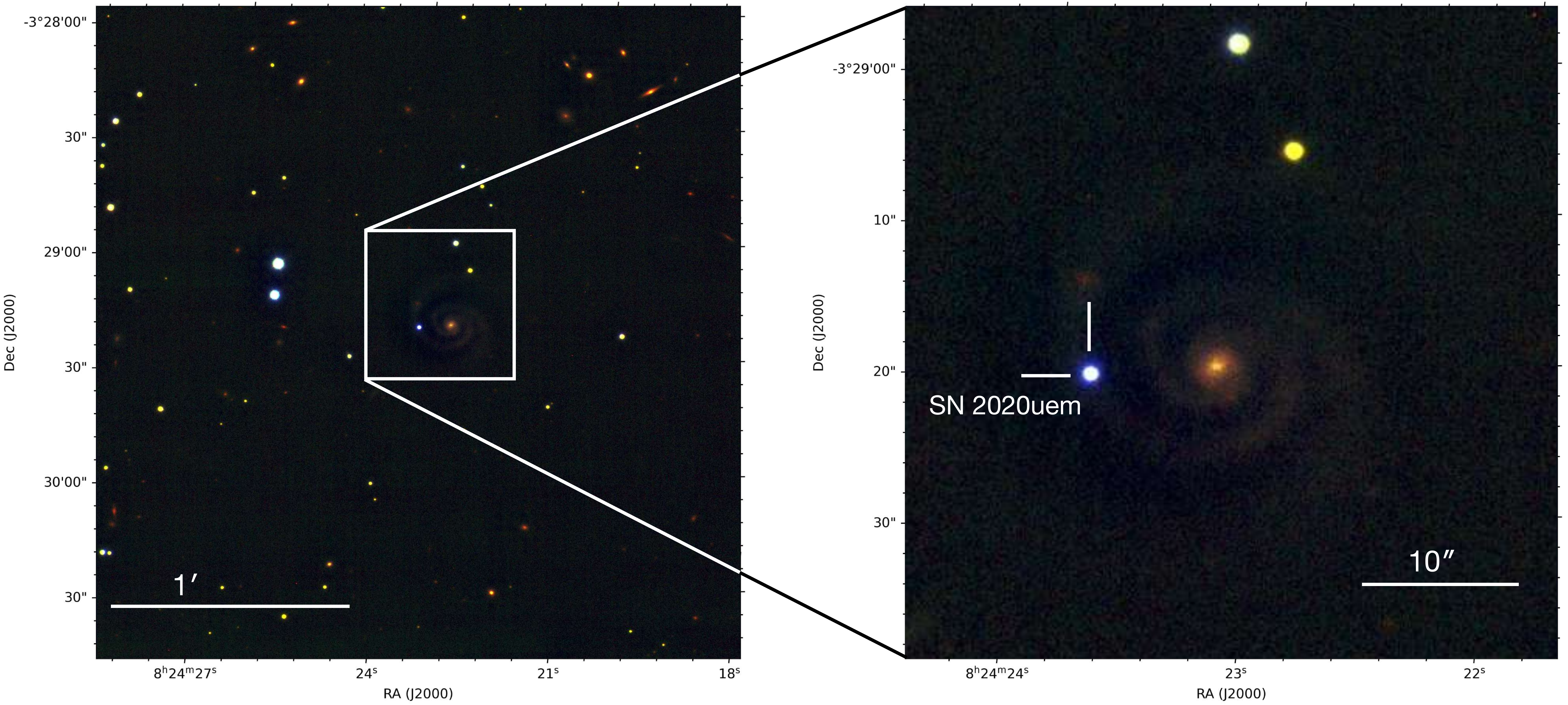}
\caption{Finding chart of SN 2020uem, from a combination of RJKs-band images obtained with the Subaru telescope on MJD 59217.20 (the R-band image taken with FOCAS) and 59311.26 (the JKs-band images taken with SWIMS).}
\label{fig:1}
\end{figure*}

SN 2020uem/ATLAS20bbsz was discovered on September 22.602 2020 UT (MJD 59114.602) by the ATLAS \citep{Tonry2020TNS}. The last non-detection of SN 2020uem was on June 15.437 2020 UT (MJD 59015.437)  with an upper limit of $21.5$ mag by the Gaia satellite \citep{Gaia2016AA}. The coordinate is $\rm{RA(J2000.0)} = 08^{\rm h}24^{\rm m}23^{\rm s}.85$ and ${\rm Dec(J2000.0)} = -03^{\circ}29^{\prime}19^{\prime\prime}.1$. SN 2020uem is located $7.7^{\prime\prime}$ east and $0.5^{\prime\prime}$ south of the host-galaxy core (see Figure \ref{fig:1}). After the discovery, the transient was classified as a Type Ia, peculiar Type IIn, and Type IIP SN \citep{Dahiwale2020TNS,Reguitti2020ATel,Reguitti2020TNS}. However, since the spectra and light curves are similar to those of SNe IIn/Ia-CSM (see Sections \ref{sec:3} and \ref{sec:4}), we classify it as a Type IIn/Ia-CSM SN.

From the position of the narrow $\rm H\alpha$ emission line of SN 2020uem, we estimate its redshift as $z= 0.041$. Adopting $H_{0} = 73.2 \pm 2.3 {\rm ~ km~s^{-1}~Mpc^{-1}}$ \citep{Burns2018ApJ}, $\Omega_{M} = 0.27$, and $\Omega_{\Lambda} = 0.73$, we derive the luminosity distance as $173.3 \pm 5.7 {\rm ~ Mpc}$, and thus the distance modulus as $36.19 \pm  0.07 {\rm ~ mag}$.

According to \citet{Schlafly2011ApJ} and the NASA/IPAC Extragalactic Database\footnote{\url{https://ned.ipac.caltech.edu/}} (NED), the Milky Way extinction in the direction of SN 2020uem is $A_{\rm V}^{\rm MW} = 0.131 {\rm ~mag}$. We assume the \citet{Fitzpatrick1999PASP} reddening law, which is characterized by $R_{\rm V} = 3.1$. In this paper, we assume that the host-galaxy extinction is negligible. The validity of this assumption is justified by the non-detection of \ion{Na}{1} D absorption feature and the location of the SN at the edge of a spiral arm, i.e., isolated from the host galaxy core.

\subsection{observations and reduction} \label{sec:2.2}

We performed optical/NIR photometry with the Hiroshima One-shot Wide-field Polarimeter \citep[HOWPol;][]{Kawabata2008SPIE} and the Hiroshima Optical and NearInfraRed Camera \citep[HONIR;][]{Akitaya2014SPIE} mounted on the 1.5m Kanata telescope at the Higashi-Hiroshima Observatory, Hiroshima University. We also performed NIR photometry with the Simultaneous-color Wide-field Infrared Multi-object Spectrograph \citep[SWIMS;][]{Konishi2012SPIE} on the 8.2m Subaru telescope. Besides, we obtained the data with the 3.8m Seimei telescope \citep{Kurita2020PASJ} at the Okayama Observatory, Kyoto University; we performed late-phase optical photometry with the TriColor CMOS Camera and Spectrograph (TriCCS)\footnote{\url{http://www.o.kwasan.kyoto-u.ac.jp/inst/triccs/}}. Optical spectroscopy was also performed with the Seimei telescope, equipped with the Kyoto Okayama Optical Low-dispersion Spectrograph with optical-fiber Integral Field Unit \citep[KOOLS-IFU;][]{Matsubayashi2019PASJ}, and the Subaru telescope equipped with the Faint Object Camera and Spectrograph \citep[FOCAS;][]{Kashikawa2002PASJ} and the High Dispersion Spectrograph \citep[HDS;][]{Noguchi2002PASJ}. NIR spectroscopy is performed with SWIMS. Note that we also performed imaging polarimetry and spectropolarimetry (see Paper II).

\subsubsection{Photometric Observations} \label{sec:2.2.1}

We obtained BVRI-band images in 19 nights with HOWPol and gri-band images in 3 nights with TriCCS. We also performed JHKs-band photometry in 22 nights with HONIR and SWIMS. We follow a standard procedure for the CCD and CMOS photometry. The data reduction was performed with the DAOPHOT package for the Point Spread Function (PSF) fitting photometry in IRAF\footnote{IRAF is distributed by the National Optical Astronomy Observatory, which is operated by the Association of Universities for Research in Astronomy (AURA) under a cooperative agreement with the National Science Foundation.}. For the late-phase data of TriCCS, we performed a pseudo-host galaxy subtraction since the supernova is faint; assuming that the host galaxy has symmetric structure, we rotated the host galaxy image by 180 degrees around the center of the host galaxy, and subtracted it from the original image at the SN location. 
For the photometric data, we performed relative photometry with the comparison stars in the same frames. The optical magnitudes of the comparison stars are taken from the APASS catalog \citep{Henden2015AAS} and the Pan-STARRS catalog \citep{Chambers2016arXiv}, and the NIR magnitudes are taken from the 2MASS catalog \citep{Persson1998AJ}. In Tables \ref{tab:optical_photometry} and \ref{tab:NIR_photometry}, we list the summary of the optical and NIR magnitudes in the Vega system.

\subsubsection{Spectroscopic Observations} \label{sec:2.2.2}

We obtained optical spectra of SN 2020uem with KOOLS-IFU (see Table \ref{tab:spectroscopy}). We used the grisms of VPH-blue and VPH683. The wavelength coverages of VPH-blue and VPH683 are 4100-8900 \text{~\AA} and 5800-8000 \text{~\AA}, respectively. The wavelength resolutions $R = \lambda/\Delta\lambda$ of VPH-blue and VPH683 are $\sim 500$ and $\sim 2000$, respectively. For the wavelength calibration, we used Hg, Ne, and Xe lamps. The data reduction was performed with the Hydra package in IRAF and a reduction software specifically developed for KOOLS-IFU data\footnote{\url{http://www.o.kwasan.kyoto-u.ac.jp/inst/p-kools/reduction-201806/index.html}}.

On January 3.20 2021 UT (MJD 59217.20) and 4.15 2021 UT (MJD 59218.15), we obtained optical spectroscopic data with a high Signal-to-Noise ratio with FOCAS (see Table \ref{tab:spectroscopy}). We used the B300 grating with no filter. The wavelength coverage is $3650-8300$ {\text \AA} with the wavelength resolution of $\sim 500$. For the wavelength calibration, we used arc lamp (Th and Ar) data. Note that we combined the data obtained in two consecutive nights, assuming that the variation of the parameters in one day is negligible because the characteristic timescale must be much longer in the late phase.

We also obtained high-resolution spectroscopic data with HDS on 2021-03-26 (UT). The echelle setup is chosen to cover the wavelength range of $4400-7100$ {\text \AA} with the spectral resolution of $\sim 50000$. Three exposures of 1800 seconds each were obtained, with the total exposure time of 5400 seconds. The airmass was $\sim 1.1$. We followed standard procedures to reduce the data. The wavelength calibration was performed using Th–Ar lamps. A heliocentric velocity correction was applied to each spectrum. The sky subtraction was performed using the data at the off-target position in the target frames. Flux calibration was performed using the spectrum of HD69503 as a standard star.  

Our NIR spectroscopy was performed with SWIMS (see Table \ref{tab:spectroscopy}). We used the zJ grism in the blue channel and the HKs grism in the red channel with $0.5 ^{\prime\prime}$ slit width. This configuration allows the spectral resolution of $\sim 1000$ and the wavelength coverage of $0.9-2.5 {\rm ~\mu m}$, which is adequate to analyze possible hot dust emission components (H and Ks bands) and emission lines of our interest. For the data reduction, we performed a standard procedure for NIR spectroscopy. For the wavelength calibration, we used Th-Ar lamps, and we further checked the accuracy of the wavelength calibration using OH airglow. The telluric absorption was corrected with the standard star, FS127, which is an F9 star \citep{Hawarden2001MNRAS}. We have taken the NIR standard star with similar airmass immediately after the target observation. For the flux calibration, we used the F9 standard star spectrum assuming a blackbody function of $6050 {\rm ~K}$. Finally, the absolute flux scale was anchored with the SWIMS photometry.

\section{Photometry} \label{sec:3}

\subsection{Light Curves} \label{sec:3.1}

\begin{figure*}
\epsscale{1.17}
\plotone{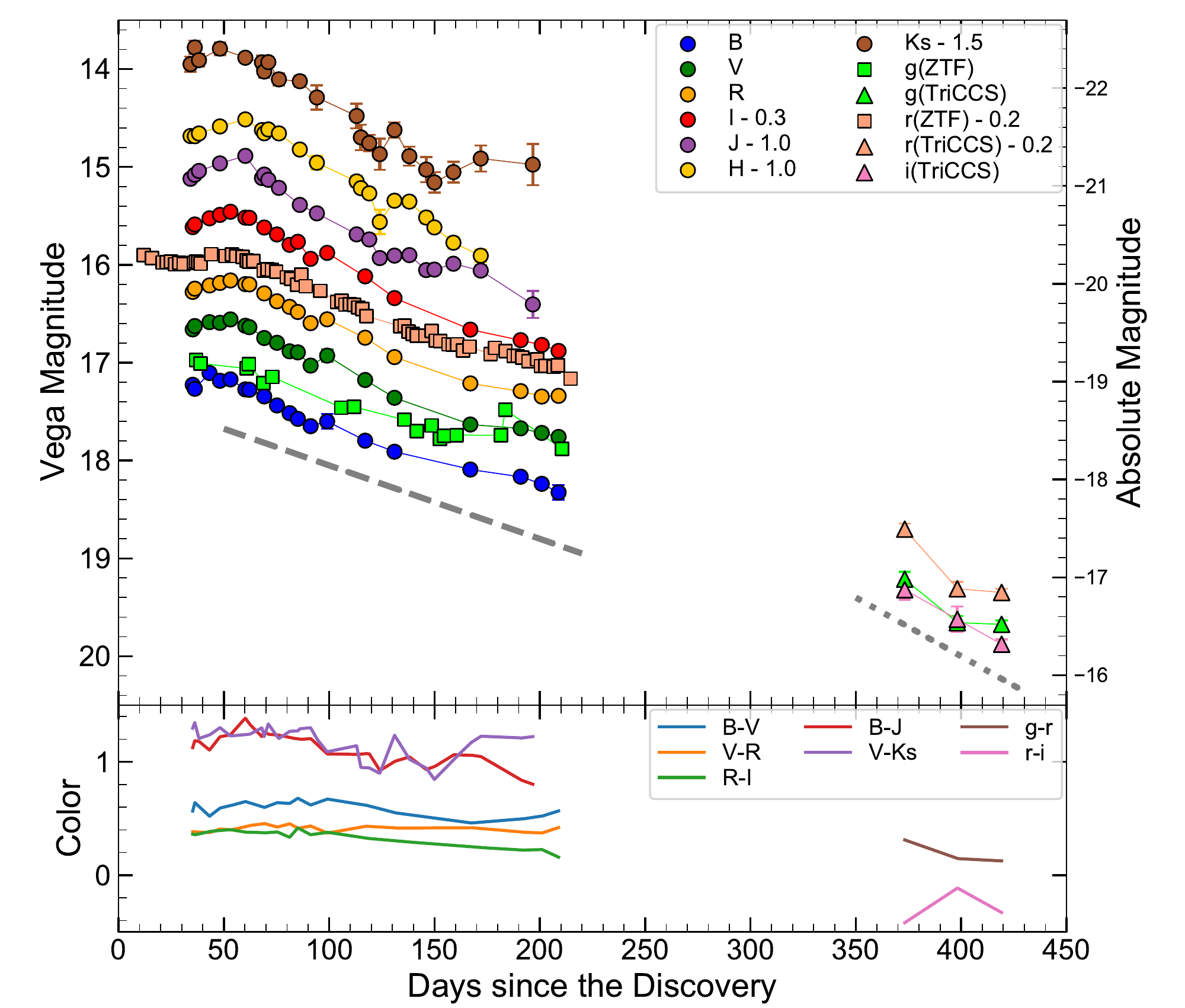}
\caption{Top Panel: Optical (BgVrRiI-band) and NIR (JHKs-band) light curves of SN 2020uem. The magnitudes in some band passes are shifted by the values shown in the legend. The left axis shows the apparent magnitude and the right axis shows the absolute magnitude, both in the Vega system. The gray dashed and dotted lines show decline rates of $0.75 {\rm ~mag}/100 {\rm ~days}$ and $1.2 {\rm ~mag}/100 {\rm ~days}$, respectively. Bottom Panel: Color evolution ($B-V$, $V-R$, $R-I$, $B-J$, $V-Ks$, $g-r$, and $r-i$) of SN 2020uem.}
\label{fig:2}
\end{figure*}

Figure \ref{fig:2} shows the light curves of SN 2020uem in optical/NIR bands. Our optical to NIR photometry has a total of 10 filters (B, g, V, r, R, i, I, J, H, Ks). Additionally, we also plot the ZTF g- and r-band data, which are available in the ZTF Public Data Release 6\footnote{\url{https://www.ztf.caltech.edu/news/dr6}}. The data cover the period from $\sim 11$ days to $\sim 419$ days after the discovery. Since the SN was discovered just after the solar conjunction, the rising phase is likely missed. All of the magnitudes are in the Vega system, and corrected for the extinction within the Milky Way. Note that the original ZTF magnitudes are described in the AB system, but we converted the magnitudes into the Vega system.

In Figure \ref{fig:2}, a bumpy structure around 50 days is seen in most of the bandpasses, which might be taken as the light curve peaks at a first look. However, the r-band ZTF light curve shows that SN 2020uem indeed showed brighter luminosity in the earlier phase and then decayed, followed by the `second' peak at $\sim 50$ days. This behavior is also seen in the ZTF g-band light curve presented by some brokers, e.g., ALeRCE \citep{Forster2021AJ}\footnote{see, e.g., \url{https://alerce.online/object/ZTF20accmutv}}. Thus, the bumpy structure around 50 days is not a real peak, and we suggest that the peak magnitude of SN 2020uem is brighter than $-19.5$ mag in the V-band. 

In the early phase (until about 50 days after the discovery), the multi-band light curves are flat or slightly increasing, and then it begins to decay with a rate of $\sim 0.75 {\rm ~mag}/100 {\rm ~days}$. The decline rate is slower than that of other Type Ia SNe and similar to interacting SNe (see also Figure \ref{fig:4}). Besides, Figure \ref{fig:2} also shows that the colors in the optical bands are almost unchanged, which indicates that the photospheric temperature keeps constant (see also Figure \ref{fig:3}). On the other hand, the Ks-band light curve shows a rebrightening after $\sim 150$ days. This might be due to thermal emission from newly-formed dust or pre-existing dust.

Although there are no observational data between $\sim 210$ days and $\sim 370$ days because the SN was again behind the sun, the optical bands likely show an accelerated decline with a rate of $\sim 1.2 {\rm ~mag}/100 {\rm ~days}$ in the late phase ($\gtrsim 350$ days). There are a few possible scenarios for this accelerated decay. One scenario is that the CSM interaction becomes weakened during this period due to the CSM density structure becoming steeper reflecting termination of intensive mass loss. In this case, we may constrain the CSM mass using the timescale of the interaction (see Paper II). Another scenario is that the optical emissions are suppressed by newly-formed dust. In fact, some cases have been reported for interacting SNe in which newly-formed dust begins to form approximately several hundred days after the explosion, and thereafter they show the NIR excess and decay in optical bands \citep[e.g.,][]{Maeda2013ApJ}. We discuss the dust formation and dust properties in Section \ref{sec:5}. Yet another scenario is due to the shock kinematics. When the shocked CSM mass becomes comparable to the ejecta mass, the shock kinematics approaches the Sedov solution. In this case, the reverse shock stops playing a role and the forward shock is slowing down rapidly, which may lead to the accelerated decay in the light curve \citep[e.g.,][]{Svirski2012ApJ, Ofek2014ApJ, Inserra2016MNRAS}.

\subsection{Quasi-Bolometric Luminosity} \label{sec:3.2}

\begin{figure}
\epsscale{1.17}
\plotone{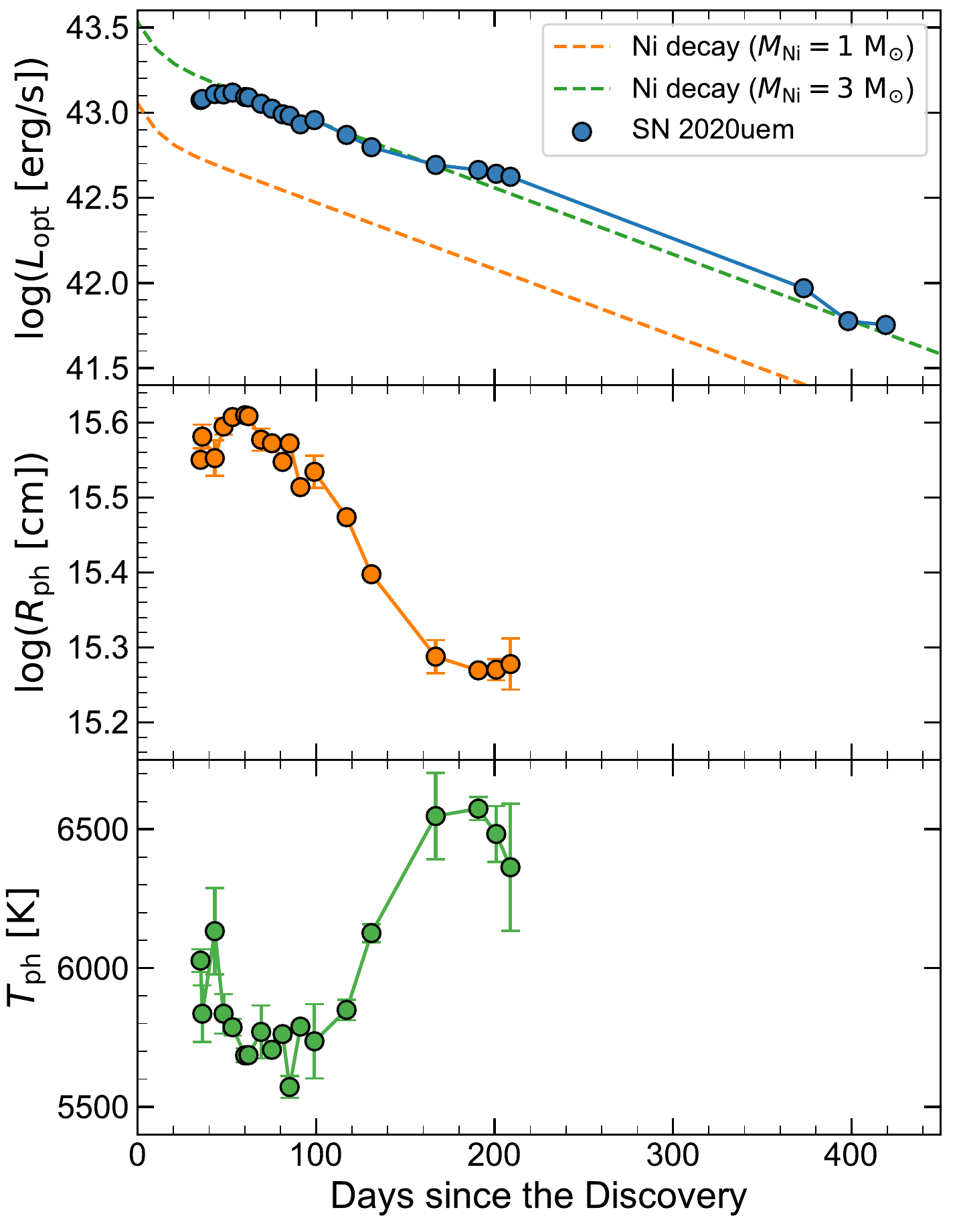}
\caption{Top: quasi-bolometric (B to I or g to i band) light curve of SN 2020uem. The orange and green dashed lines show light curves powered by the radioactive decay of $1 {\rm ~M_{\odot}}$ and $3 {\rm ~M_{\odot}}$ of $\rm ^{56}Ni$ , respectively. Middle: time evolution of the photospheric radius estimated by SED fitting. Bottom: time evolution of the photospheric temperature.}
\label{fig:3}
\end{figure}

We compute the quasi-bolometric luminosity by integrating the flux of each optical band. We perform a blackbody fit to the spectral energy distribution (SED) using the optical and NIR photometry (BVIJHKs). We do not use the R-band magnitude since it covers a strong $\rm H\alpha$ emission (see also Section \ref{sec:4}). We estimate the photospheric radius ($R_{\rm ph}$) and the photospheric temperature ($T_{\rm ph}$) of SN 2020uem. Note that the blackbody fit is not performed for the late-phase data, because we have data only in 3 bands, including the r band that also covers a strong $\rm H\alpha$ emission. Figure \ref{fig:3} shows the time evolution of the quasi-bolometric luminosity, photospheric radii, and temperatures.

The maximum optical luminosity after the discovery is over $10^{43} {\rm ~erg~s^{-1}}$. In Figure \ref{fig:3}, we also plot the light curves powered by radioactive decay of $^{56}{\rm Ni}$ and $^{56}{\rm Co}$. To explain the high luminosity, we need $^{56}{\rm Ni}$ mass of $\sim 3 {\rm ~M_\odot}$, which is beyond the feasible regime of SNe, including both thermonuclear and core-collapse SNe \citep[e.g.,][]{Yamanaka2009ApJ, Ouchi2021ApJ}. The result supports that the SN is powered by the CSM interaction. The estimated photospheric radii and temperatures are roughly $\sim 10^{15} {\rm ~cm}$ and $\sim 6000 {\rm ~K}$, respectively. These are consistent with other interacting SNe \citep[e.g.,][]{Taddia2020AA,Moriya2020AA}. In particular, the roughly constant temperature of $\sim 6000 {\rm ~K}$ may suggest that the photosphere is related to a hydrogen recombination front.

\begin{figure}
\epsscale{1.17}
\plotone{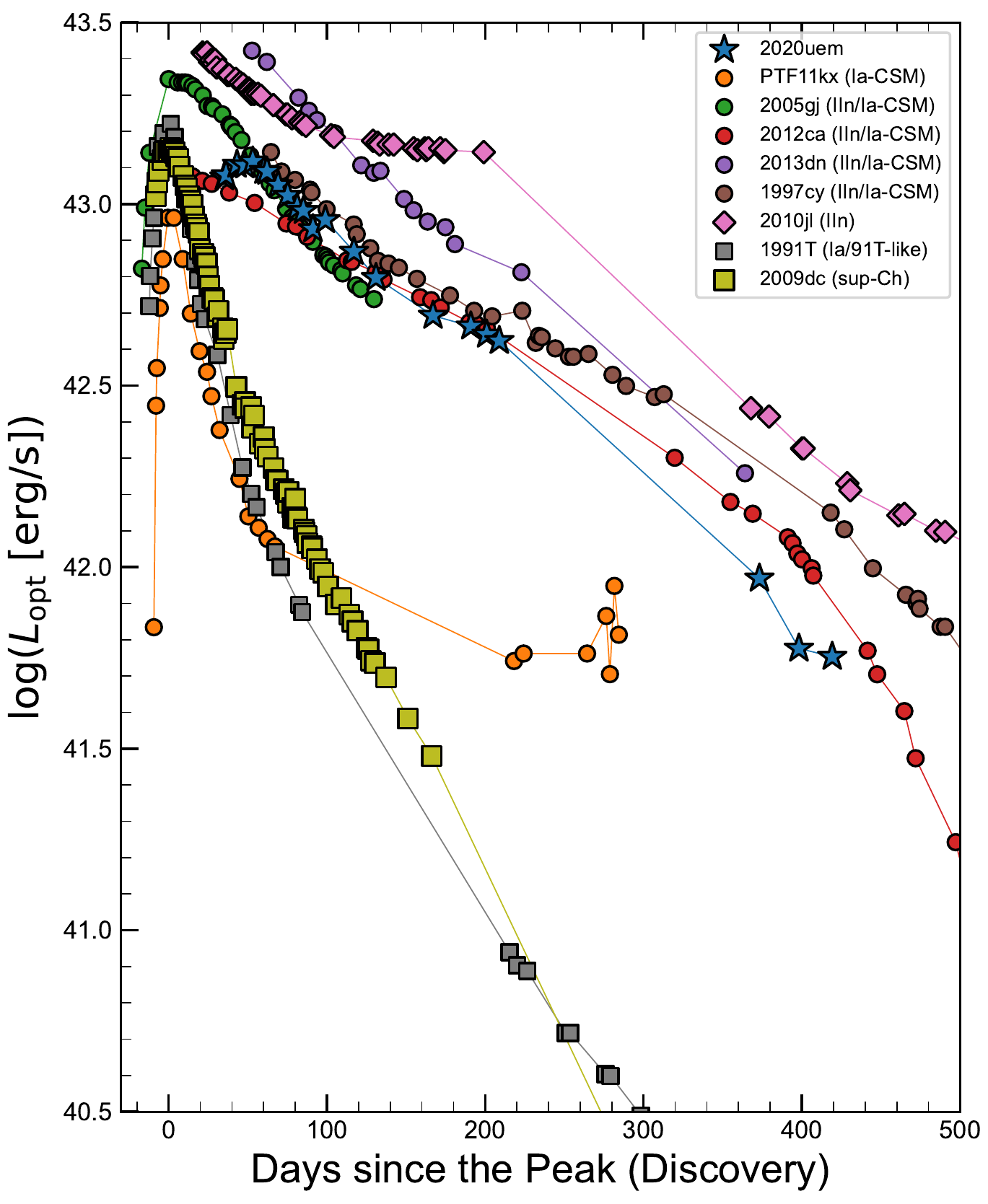}
\caption{Optical bolometric light curves of SN 2020uem and other SNe, including Type Ia-CSM; PTF11kx \citep{2013ApJ...772..125S}, Type IIn/Ia-CSM; SN 1997cy \citep{Germany2000ApJ,Turatto2000ApJ}, SN 2005gj \citep{Prieto2007arxiv}, SN 2012ca \citep{Inserra2014MNRAS, Inserra2016MNRAS}, SN 2013dn \citep{Fox2015MNRAS}, Type Ia/91T-like; SN 1991T \citep{1998AJ....115..234L}, super-Chandrasekhar Type Ia; SN 2009dc \citep{Yamanaka2009ApJ}, and Type IIn; SN 2010jl \citep{Zhang2012AJ}. Note that the epochs of all light curves, except for SN 2020uem, are relative to the peak, while the epoch of SN 2020uem is relative to the discovery because of the lack of observations in the rising phase.}
\label{fig:4}
\end{figure}

Figure \ref{fig:4} shows a comparison of the bolometric light curve of 2020uem with those of other SNe. Although we have uncertainties in the explosion date, it is clear that SN 2020uem is as bright as other SNe Ia-CSM and SNe IIn/Ia-CSM and the timescale of the light curve evolution is also similar to other interacting SNe. On the other hand, the luminosity of SN 2020uem is brighter than those of luminous non-interacting SNe Ia, e.g., SN 1991T and SN 2009dc. The post-peak light curve of SNe Ia is powered by radioactive decay, but the decline rate of SN 2020uem is much slower than that of SNe Ia. This supports that the interaction is the main energy source of SN 2020uem (see also Section \ref{sec:6} and Paper II).

In the late phase ($\gtrsim 300$ days), interacting SNe generally show an accelerated decay. The decline rate of SN 2020uem is similar to other interacting SNe: SN 2013dn, SN 1997cy, and SN 2010jl. The late-phase fading can be in part attributed to newly-formed dust \citep{Maeda2013ApJ}, while the origin of the fading is generally not clear.

\section{Spectroscopy} \label{sec:4}

\subsection{Spectral Evolution} \label{sec:4.1}

\begin{figure}
\epsscale{1.17}
\plotone{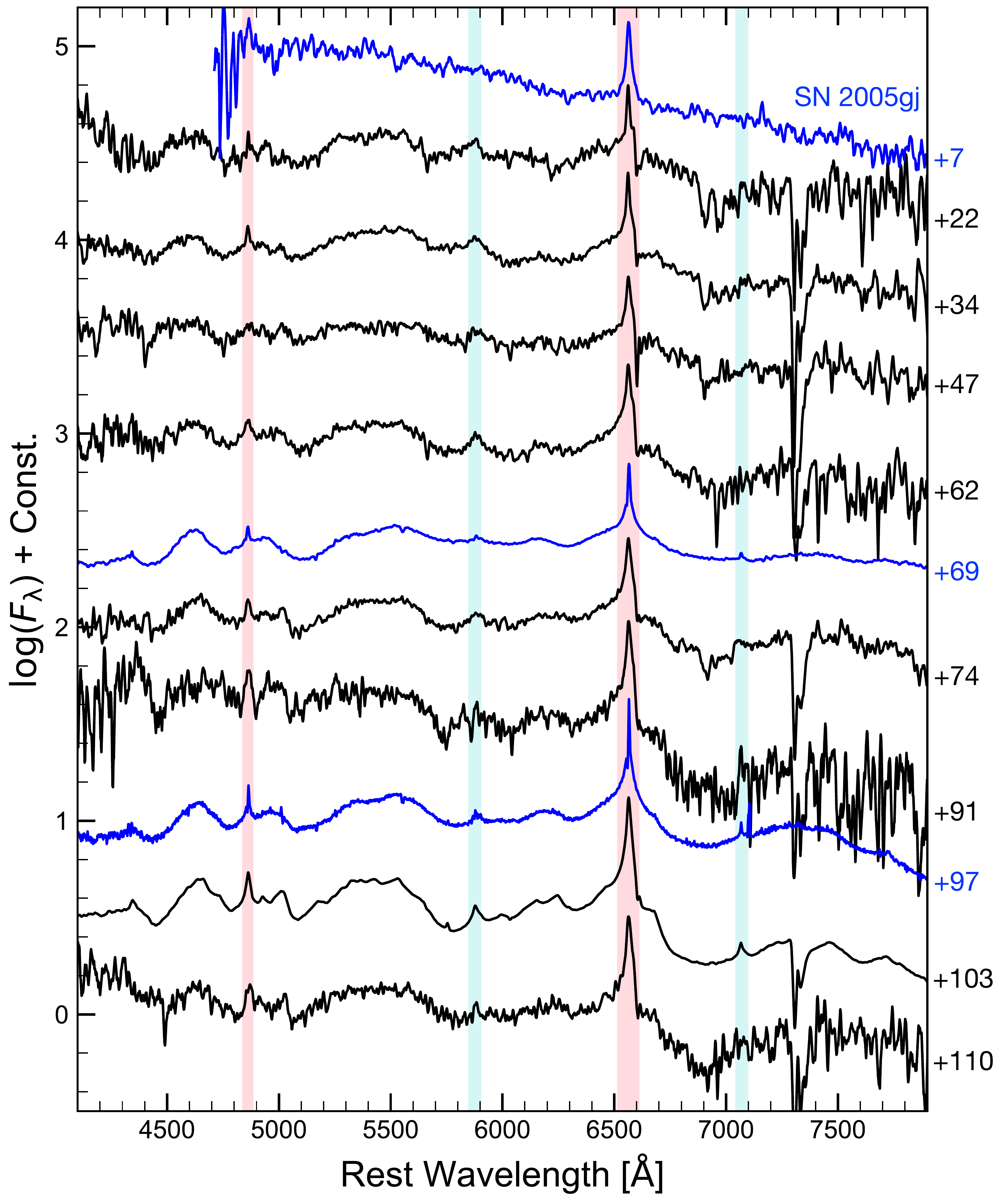}
\caption{Spectral evolution of SN 2020uem (black lines) extending from $+22$ days to $+110$ days after the discovery. The epoch for each spectrum is shown on the right side. All spectra are corrected to the rest wavelength with the redshift $z =  0.041$. The red shaded regions are the wavelengths of $\rm H\alpha$ and $\rm H\beta$. The blue shaded regions are the wavelengths of \ion{He}{1} $\lambda 5876$ and \ion{He}{1} $\lambda 7065$. For comparison, we plot the spectra of an SN IIn/Ia-CSM, SN 2005gj \citep{Aldering2006ApJ, Silverman2012MNRAS, Silverman2013ApJS}, with blue lines.}
\label{fig:5}
\end{figure}

Figure \ref{fig:5} shows the optical spectra of SN 2020uem, together with those of SN IIn/Ia-CSM 2005gj for comparison. The spectra of SN 2005gj are gathered from the Weizmann Interactive Supernova data Repository\footnote{\url{https://www.wiserep.org/}} \citep[WISeRep;][]{Yaron2012PASP}. SN 2020uem shows a prominent narrow $\rm H\alpha$ emission line in all spectra, and also exhibits other Balmer lines and \ion{He}{1} lines in good-quality spectra. These narrow emission lines suggest that the SN ejecta are colliding with a substantial amount of an H-rich CSM. Besides, the spectra also show broad features, e.g., features around $\sim 4650$ {\text \AA} and $\sim 5500$ {\text \AA}. The spectral features in the wavelengths bluer than $\sim 5500$ {\text \AA} are likely produced by overlapping emission lines of Fe-group elements, i.e., `quasi-continuum' \citep[e.g.,][]{Silverman2013ApJS,Inserra2014MNRAS, Fox2015MNRAS}. These spectral features do not evolve significantly over $\sim 100$ days. This result suggests that the photosphere and line-forming regions do not change, and emission lines originating from the underlying SN may escape from the CSM in the diffusion time. Therefore, the non-evolving features also support that the CSM interaction is still dominant during the period. We identify SN 2020uem as an SN IIn/Ia-CSM based on these spectral features in addition to the light-curve evolution. Note that we rule out a possibility that SN 2020uem is powered by a CSM interaction with SN Ic ejecta (see also Section \ref{sec:6.3}).

\subsection{Balmer line profiles} \label{sec:4.2}

\begin{figure}
\epsscale{1.17}
\plotone{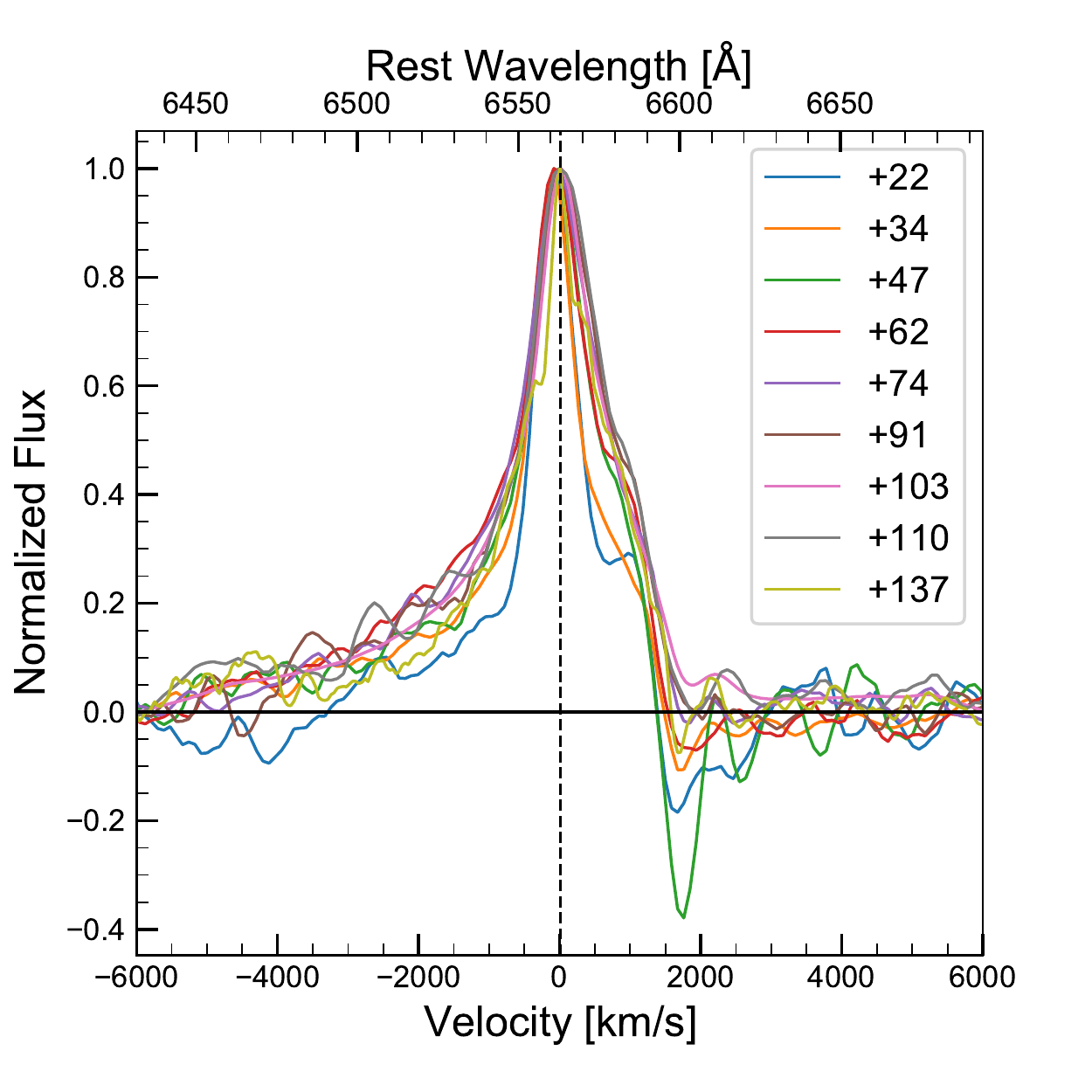}
\caption{Time evolution of the $\rm H\alpha$ profile, plotted in velocity space. We also show the rest wavelength in the top axis. The spectra are scaled by the maximum flux of $\rm H\alpha$ and the continuum is already subtracted. The absorption at $v \approx 2000{\rm ~km~s^{-1}}$ of low quality spectra may be oversubtraction of a nitrogen sky emission line.}
\label{fig:6}
\end{figure}

Figure \ref{fig:6} shows the time evolution of the $\rm H\alpha$ line. The line profile does not change over $130$ days. The profile shows a sharp peak around its rest wavelength ($6563$ {\text \AA}) with excess in bluer wavelengths. In general, SNe IIn/Ia-CSM have an asymmetric $\rm H \alpha$ profile, which shows an excess in the blueshifted side \citep[e.g.,][]{Deng2004ApJ,Dilday2012Science,Inserra2016MNRAS}. On the other hand, the line profile of $\rm H\beta$ is symmetric. NIR hydrogen lines also show symmetric profiles (see Figure \ref{fig:9}). Based on these considerations, the asymmetric structure in the $\rm H\alpha$ profile most likely originates from contamination by other lines. In some spectral synthesis calculations for SNe Ia \citep[e.g.,][]{Childress2015MNRAS}, forbidden iron lines ([\ion{Fe}{2}]) are expected to be dominant around $\sim 6500$ {\text \AA} at $\sim 100$ days after the explosions. Indeed, SNe IIn/Ia-CSM show broad features potentially associated with [\ion{Fe}{2}] and [\ion{Fe}{3}] (see also Sections \ref{sec:4.4} and \ref{sec:6.3}). Therefore, we propose the asymmetry of the $\rm H\alpha$ profile is due to the contamination of the broad iron lines. This result also may support the picture that the underlying SN behind SNe IIn/Ia-CSM is related to SNe Ia. Note that the detailed radiation mechanisms in SNe IIn/Ia-CSM still remain unclear. In general, normal SNe Ia are much dimmer than the continuum components seen in SNe IIn/Ia-CSM, and thus it may be difficult for the SN Ia component to affect the emission profile. However, in the interacting SNe, the underlying SN ejecta are illuminated by the high-energy photons from the CSM interaction, and thus the thermal balance there is controlled by the energy input by the interaction \citep{maeda2022}; it might lead to strong/detectable SN component in the spectra. 
As yet another possibility, the continuum may be entirely formed within the shocked CSM; \citet{Chevalier2003LNP} suggest that interacting SNe can form Fe emission in the shocked CSM, and it can be possible to explain the Fe-like contamination in the $\rm H\alpha$ profile. 
Differentiating these scenarios will require further investigation on the radiation mechanisms of interacting SNe, which is beyond the scope of the present work. 

The narrow component excluding the blue excess is well fitted by a Lorentzian function with a full-width half maximum (FWHM) velocity of $\sim 1500 {\rm ~km~s^{-1}}$. Note that the top of the narrow component is not completely resolved, which is limited by the spectral resolution (see below). In canonical (non-interacting) SNe, an FWHM is believed to be a good tracer for shock velocity. However, the shock velocity of $\sim 1500 {\rm ~km~s^{-1}}$ is inconsistent with the energy balance; assuming an SN Ia explosion, the energy release associated with the deceleration of the ejecta from $\sim 10000 {\rm ~km~s^{-1}}$ to $\sim 1000 {\rm ~km~s^{-1}}$ should be $\sim 10^{51}{\rm ~erg}$. Although the energy conversion efficiency is undefined, the released energy exceeds the total radiation energy of SN 2020uem ($\sim 10^{50} {\rm ~erg}$), which we will discuss further in Paper II. Besides, it is clear that the slow evolution of the line width does not trace the deceleration of the shock velocity required in the SN-CSM interaction scenario. Moreover, \citet{Chugai2001MNRAS} suggests that the ${\rm H\alpha}$ profiles of SNe IIn are broadened by multiple electron scatterings of the photons in the dense CSM. In any case, the mechanisms of line formation for interacting SNe still remain unclear. It is beyond the scope of this paper to discuss the detailed mechanism of the line formation.

\begin{figure}
\epsscale{1.17}
\plotone{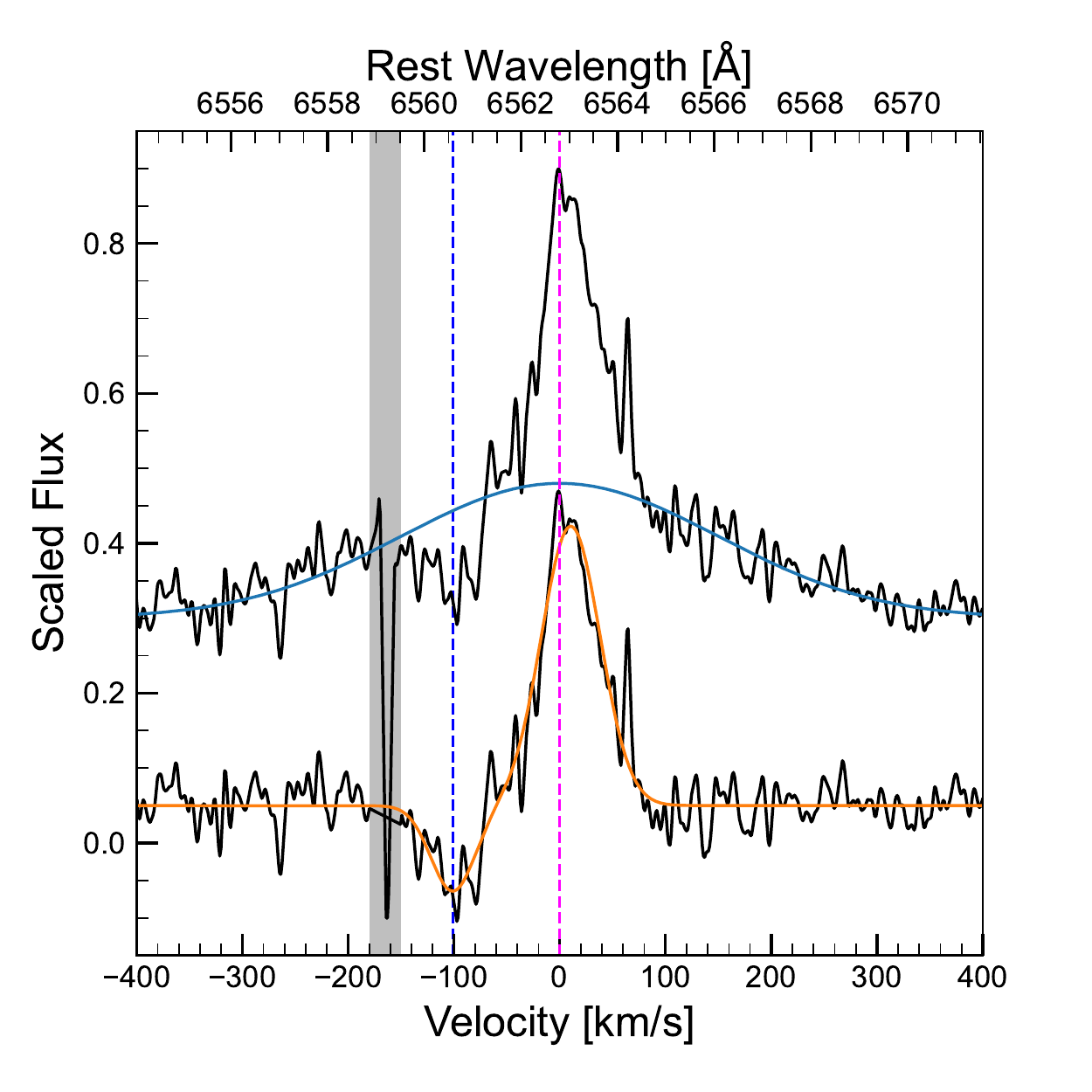}
\caption{A high dispersion spectrum of top of the narrow $\rm H\alpha$ profile, plotted in velocity space and arbitrary flux scale. We also show the rest wavelength in the top axis. The upper black spectrum shows a non-continuum-subtracted spectrum, while the lower black spectrum is continuum-subtracted spectrum. The blue solid line shows an underlying broad component fit by a single Gaussian function. The solid orange line shows the double Gaussian components to fit the narrow P-Cygni component. The magenta vertical dashed line shows the rest wavelength of $\rm H\alpha$ ($6562.8$ {\text \AA}). The cyan vertical dashed line shows the blueshift of $-100{\rm ~km~s^{-1}}$ from the rest wavelength, which corresponds to the absorption minimum of the P-Cygni profile. The apparent absorption at $v \approx -160 {\rm ~km~s^{-1}}$, which is shaded by gray color, is an artifact, caused by bad pixels.}
\label{fig:7}
\end{figure}

In Figure \ref{fig:7}, we show the high-dispersion spectrum taken with HDS. On the bottom, we show the one subtracted by its continuum, which is evaluated by a single Gaussian fitting. This spectrum resolves the top of the narrow $\rm H\alpha$ profile. The spectrum shows a narrow P-Cygni profile. The emission component shows an FWHM velocity of $\sim 100{\rm ~km~s^{-1}}$ and the absorption component shows a blueshift of $\sim 100 {\rm ~km~s^{-1}}$, as obtained by line fitting with double Gaussian components. The narrow P-Cygni profile is likely produced in the unshocked CSM and the velocity corresponds to the unshocked wind velocity. The wind velocity of SNe IIn/Ia-CSM observed so far is in the range of $65-300 {\rm ~km~s^{-1}}$ \citep{Inserra2016MNRAS}, e.g., $\sim 65$ km s$^{-1}$ seen in a `genuine' SN Ia-CSM PTF11kx. The CSM velocity of SN 2020uem ($\sim 100 {\rm ~km~s^{-1}}$) is consistent with the previous studies. The wind velocity of $\sim 100 {\rm ~km~s^{-1}}$ is faster than the escape velocity of a red supergiant, which may indicate that the CSM might have been formed by a dynamical process, rather than a static mass loss process. Such a dynamical mass-loss process may, for example, be realized in a merger of a white dwarf and an asymptotic giant blanch or red supergiant \citep{Livio2003ApJ, Kashi2011MNRAS, Jerkstrand2020Science}. We discuss possible progenitor scenarios further in Paper II.

SN 2020uem shows a high $\rm H\alpha / H\beta$ ratio, which is over $\sim 7$. SNe IIn/Ia-CSM are characterized by a higher $\rm H\alpha / H\beta$ ratio than those of SNe IIn. The ratio is a tracer of the CSM density. The result indicates that SNe IIn/Ia-CSM have denser CSMs than SNe IIn. We discuss the ratios in comparison with other interacting SNe in Section \ref{sec:6.2}.

\subsection{NIR Spectroscopy} \label{sec:4.3}

\begin{figure}
\epsscale{1.17}
\plotone{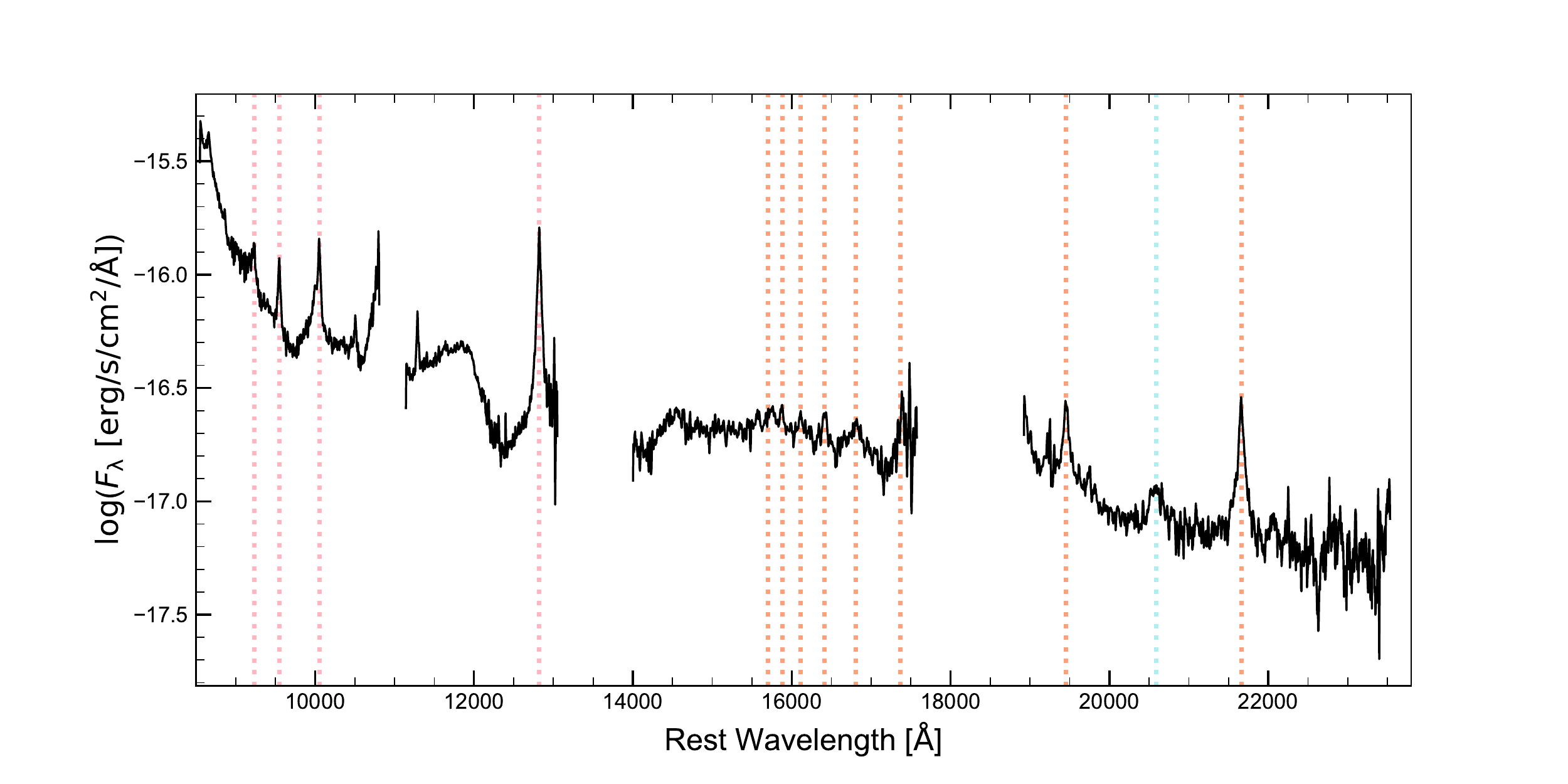}
\caption{NIR spectrum of SN 2020uem taken with SWIMS at 196.7 days after the discovery}. The spectrum contains 4 spectra in z, J, H, and Ks bands. The spectrum is corrected for the redshift and the Milky Way extinction. The flux scale is also calibrated with the J- and Ks-band photometry taken in the same night. Paschen and Brackett series are highlighted with red and orange vertical dotted lines, while \ion{He}{1} is shown in blue.
\label{fig:8}
\end{figure}

Figure \ref{fig:8} shows the SWIMS spectrum at 196.7 days after the discovery. A striking feature in our NIR spectrum is the narrow emission lines of Paschen ($\rm Pa\beta$, $\rm Pa\delta$, $\rm Pa\epsilon$) and Brackett ($\rm Br\gamma$, $\rm Br\delta$, $\rm Br\epsilon$, ...) series and \ion{He}{1}. The narrow emission lines indicate the strong CSM interaction. Another characteristic feature is a bump component around $\sim 12000$ {\text \AA}. The broad component is possibly composed of \ion{Mg}{1} $\lambda 1.183 {\rm ~\mu m}$, which may be contaminated with [\ion{Fe}{2}] lines. This is a common feature among SNe IIn/Ia-CSM \cite[e.g.,][]{Fox2015MNRAS,Inserra2016MNRAS}. In addition to the emission lines, the spectrum shows an excess at redder wavelengths. In Section \ref{sec:5}, we discuss the properties of the dust grains probably associated with this feature. 

\subsection{Paschen and Brackett line profiles} \label{sec:4.4}

\begin{figure}
\epsscale{1.17}
\plotone{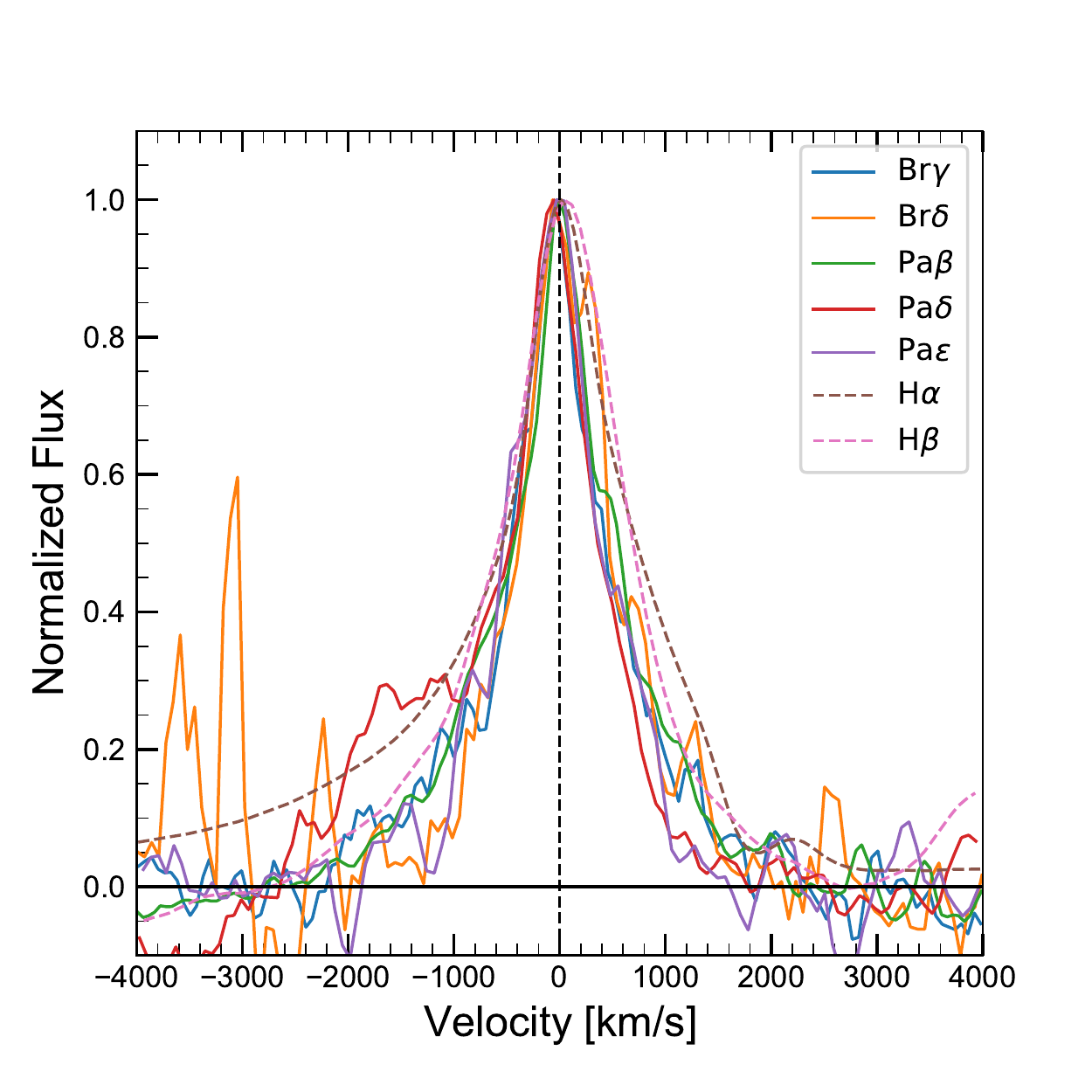}
\caption{Comparison of the line profiles, shown for $\rm Br \gamma$ (blue), $\rm Br \delta$ (orange), $\rm Pa \beta$ (green), $\rm Pa \delta$ (red), and $\rm Pa \varepsilon$ (purple). For comparison, we also plot $\rm H\alpha$ and $\rm H\beta$ obtained at $t=+103$ days with FOCAS. Each profile is normalized by its peak flux. The vertical dashed line shows the rest wavelength, i.e. $v=0 {\rm ~km~s^{-1}}$, and the horizontal line shows the zero flux level.}
\label{fig:9}
\end{figure}

Figure \ref{fig:9} shows the comparison of the Paschen and Brackett lines in the velocity space. The NIR hydrogen emission profiles are also well fitted by a narrow Lorentzian component with an FWHM velocity of $\sim 1000 {\rm ~km~s^{-1}}$. All the lines show similar symmetric profiles. Note that the $\rm H\alpha$ profile of SN 2020uem also exhibits a narrow component but with an excess in the blue side due to probable iron contamination. In fact, the symmetric profiles in the NIR wavelengths also support that the asymmetry in $\rm H\alpha$ is not likely due to the geometry or kinematics, but the contamination by other (i.e., Fe) lines.

\section{IR Excess and Dust Properties} \label{sec:5}

In this section, we present the results of the dust-emission modeling for SN 2020uem. We interpret the NIR excess as a thermal continuum from dust grains. Dust properties (dust mass, temperature, and species) provide us with insights into the nature of SNe. 

\subsection{Mass and Temperature of Dust Grains} \label{sec:5.1}

\begin{figure}
\epsscale{1.17}
\plotone{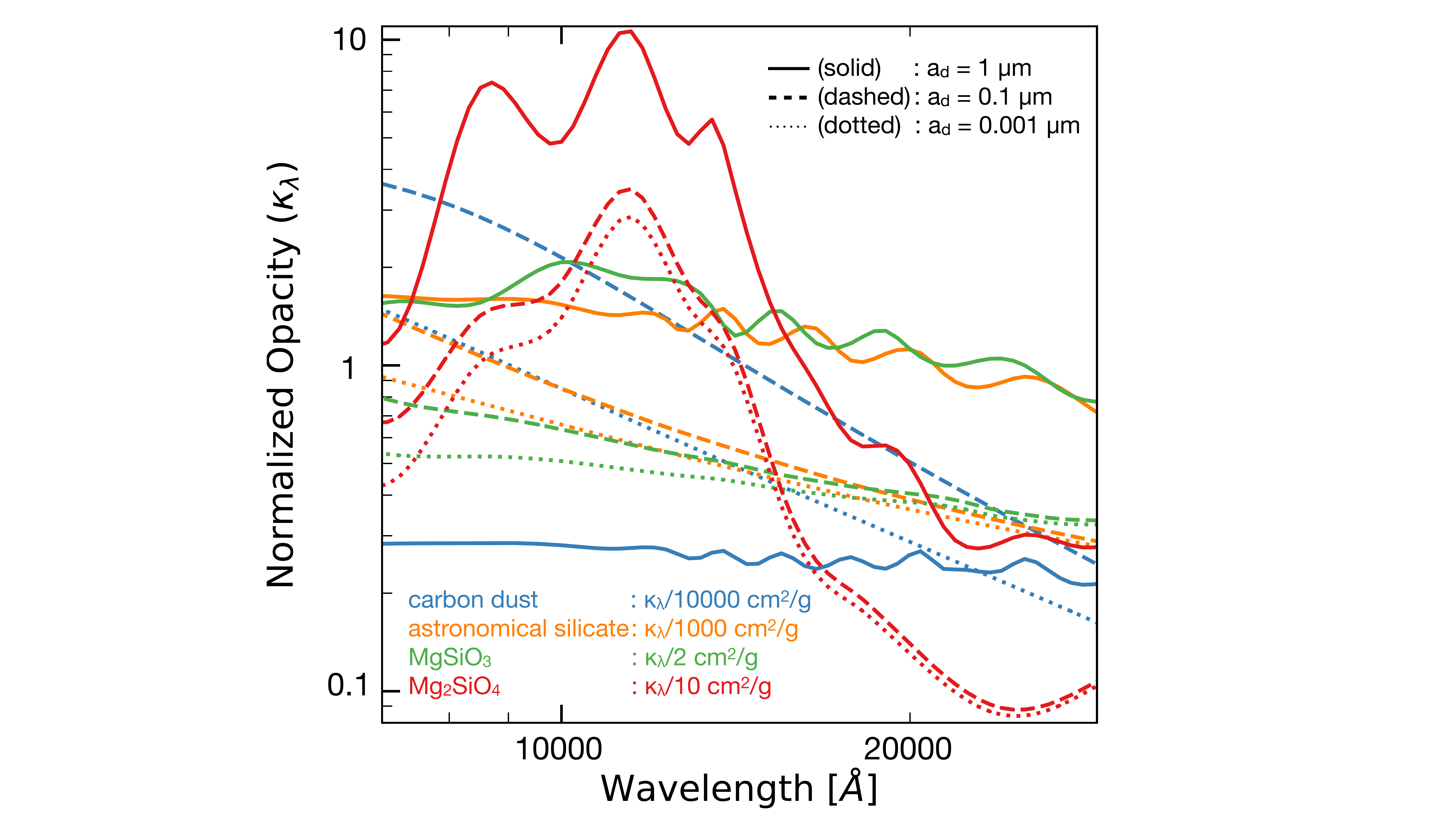}
\caption{Normalized opacity of dust species; carbon dust (blue), astronomical silicate (orange), $\rm MgSiO_{3}$ (green), and $\rm Mg_{2}SiO_{4}$ (red). The normalization constants are shown in the figure. The line style specifies the dust particle size; the solid, dashed and dotted lines are for $a_{\rm d} = 1 {\rm ~\mu m}$, $0.1 {\rm ~\mu m}$, and $0.01 {\rm ~\mu m}$, respectively.}
\label{fig:10}
\end{figure}

We perform gray-body fitting for the NIR continuum of SN 2020uem. The total mass of the dust grains ($M_{\rm d}$) and the dust temperature ($T_{\rm d}$) are connected as follows:
\begin{equation}
    F_{{\rm \nu,dust}}=\frac{\kappa_{{\rm a}, \nu} M_{\rm{d}} B_{\rm \nu}\left(T_{\rm{d}}\right)}{D^{2}},
\end{equation}
where $F_{{\rm \nu,dust}}$ is the flux of the dust thermal emission, $B_{\rm \nu}(T)$ is the blackbody function for a temperature $T$, $\kappa_{{\rm a},\nu}$ is a mass absorption coefficient of the dust grains at frequency $\nu$, and $D$ is the luminosity distance to the SN ($173.3 {\rm ~Mpc}$). In addition to the dust emission, we adopt a blackbody emission of $T_{\rm ph} \approx 6300 {\rm ~K}$ for the SN radiation, and a photospheric radius of $R_{\rm ph} \approx 1.8 \times 10 ^{15}{\rm cm}$. These values are derived by the SED fitting in the optical bandpasses (see Figure \ref{fig:3}). The total flux of the SN ($F_{\nu, \rm tot}$) is given as follows:
\begin{equation}
    F_{{\nu, \rm tot}} = \pi B_{\rm \nu, BB}(T_{\rm ph}) \left(\frac{R_{\rm ph}}{D}\right)^{2} + F_{\rm \nu,dust}.
\end{equation}
For the modeling, we adopt four dust species; carbon (graphite), astronomical silicate, $\rm MgSiO_{3}$, and $\rm Mg_{2}SiO_{4}$. We calculate the mass absorption coefficients of these dust species using optical constants by \citet{Zubko1996MNRAS} for carbon dust, \citet{Draine2003ApJ} for astronomical silicate, \citet{Dorschne1995AA} for $\rm MgSiO_{3}$, and \citet{Dorschne1995AA} for $\rm Mg_{2}SiO_{4}$. Figure \ref{fig:10} shows the opacity of these dust species as a function of wavelength.

\subsection{Results} \label{sec:5.2}

\subsubsection{Dust Properties} \label{sec:5.2.1}

\begin{figure*}
\epsscale{1.17}
\gridline{\fig{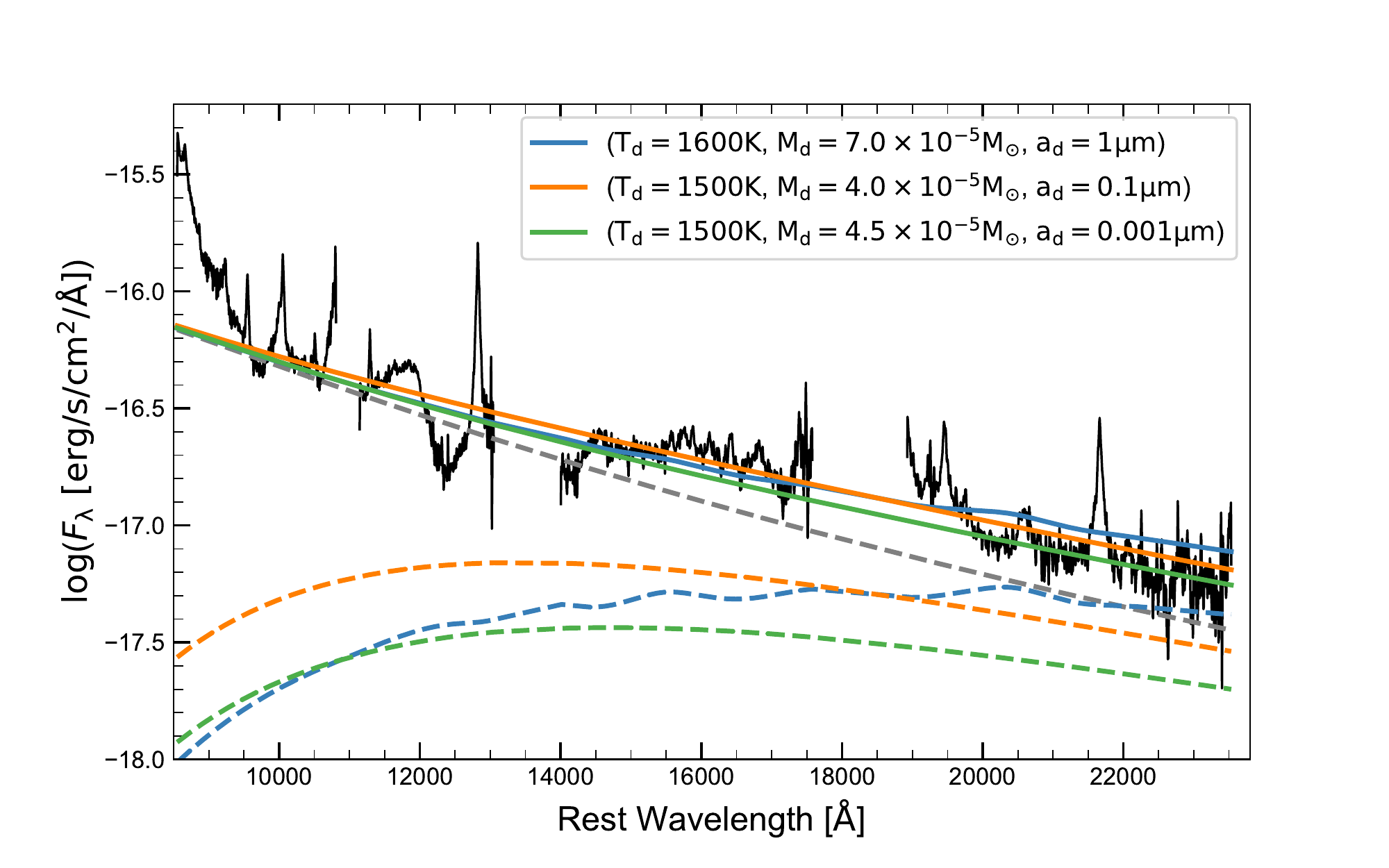}{0.45\textwidth}{(A) carbon dust}
          \fig{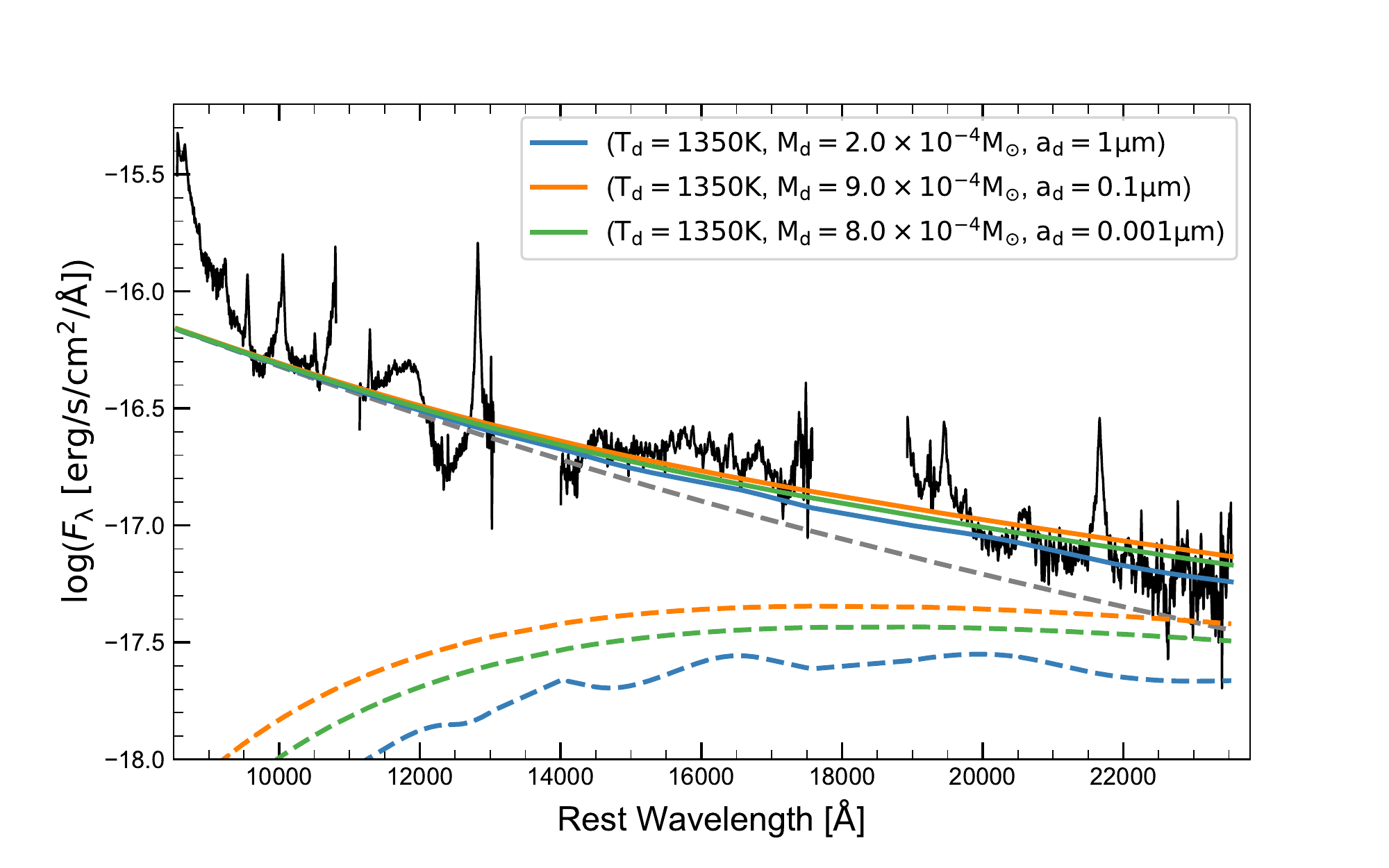}{0.45\textwidth}{(B) astronomical silicate}
          }
\gridline{\fig{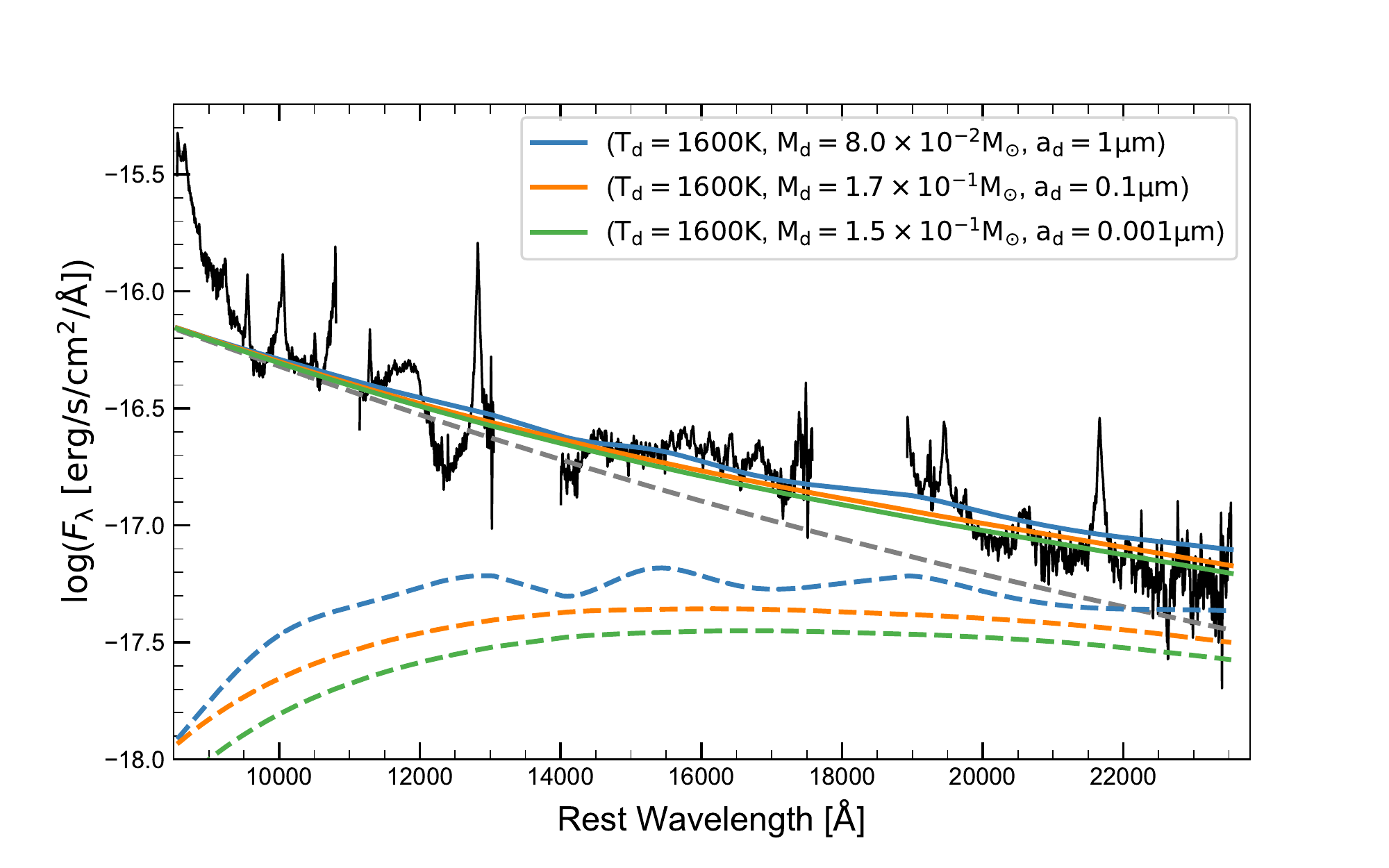}{0.45\textwidth}{(C) $\rm MgSiO_{3}$}
          \fig{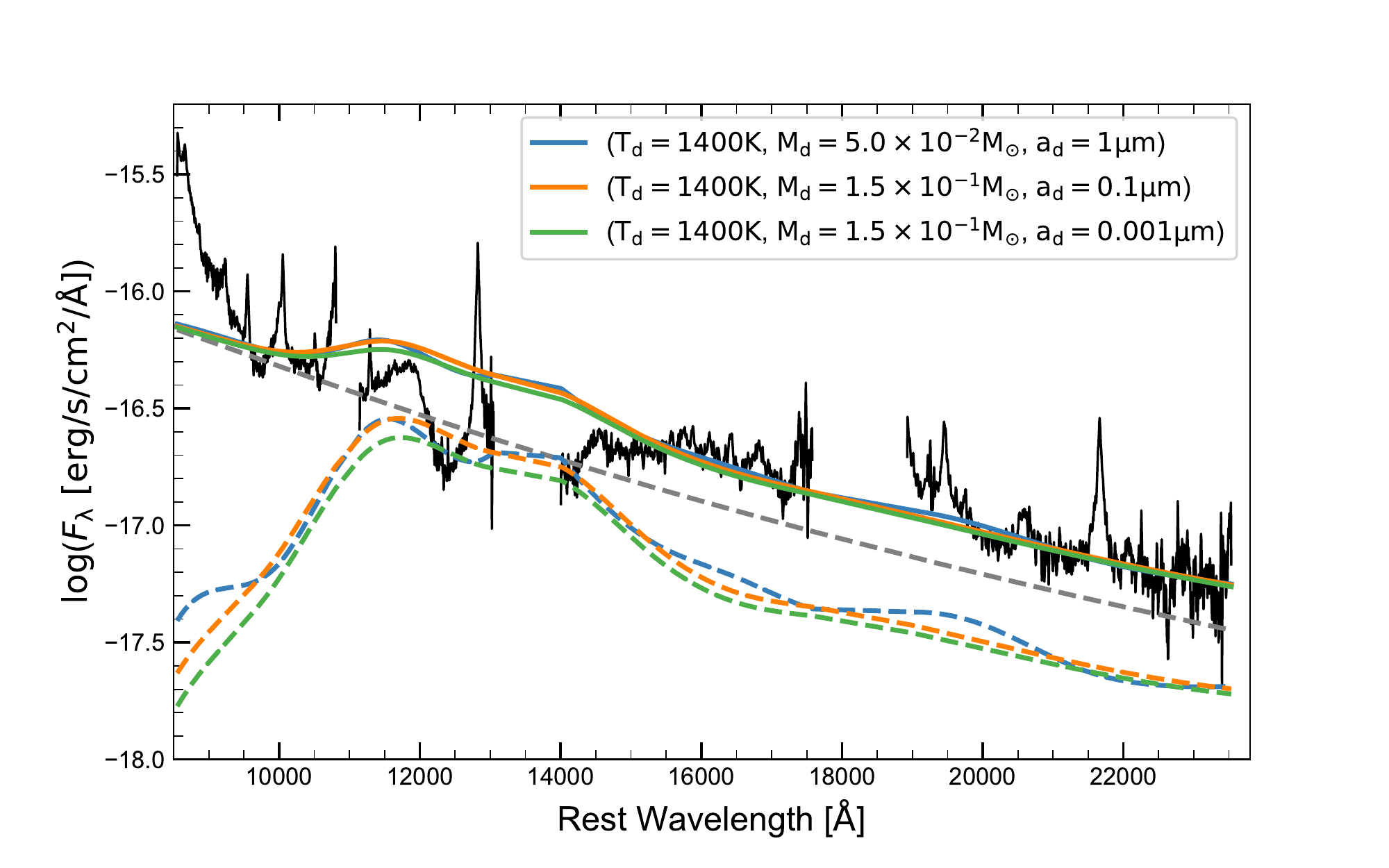}{0.45\textwidth}{(D) $\rm Mg_{2}SiO_{4}$}
          }
\caption{The observed spectrum as compared to the dust emission models for (A) carbon dust, (B) astronomical silicate, (C) $\rm MgSiO_{3}$, and (D) $\rm Mg_{2}SiO_{4}$. The gray dashed line shows blackbody radiation with a temperature of $6500 {\rm ~K}$ from the SN. The blue, orange, and green dashed lines show the dust thermal emission ($F_{\nu, \rm dust}$). The solid lines show the total flux of the dust emission model ($F_{\nu, \rm tot}$). The parameter sets of dust properties (dust temperature, mass, and size) are shown in the legends.}
\label{fig:11}
\end{figure*}

We plot comparisons between the observed spectrum with the dust emission models in Figure \ref{fig:11}. Compared to the single blackbody component of $\sim 6500 {\rm ~K}$, the thermal component including dust emission fits better the NIR continuum, especially in the H and K bands. At the wavelength range of $\lesssim 9500$ {\text \AA}, the observed spectrum is apparently brighter than the models, but this may be due to the NIR calcium emission features.

In general, assuming that the size of dust ($a_{\rm d}$) is smaller than $\sim 0.1 {\rm ~\mu m}$, the parameter set of ($T_{\rm d}$, $M_{\rm d}$) is not much affected by the dust size. On the other hand, assuming the large dust ($a_{\rm d} \gtrsim 1 {\rm ~\mu m}$), the parameter set is slightly different from the smaller ones. This is because the dust size becomes comparable to the observed wavelength in the latter case.

The fitting by $\rm MgSiO_{3}$ and $\rm Mg_{2}SiO_{4}$ grains requires a large amount of dust, i.e., $M_{\rm d} \sim 1 \times 10^{-1}{\rm ~M_{\odot}}$. This is probably too large to be realized in interacting SNe. 
Besides, the fitting by $\rm Mg_{2}SiO_{4}$ shows a bump in the J band, which does not fit the observed spectrum. Therefore, we exclude the possibility that dust is composed of $\rm MgSiO_{3}$ and/or $\rm Mg_{2}SiO_{4}$.

For carbon dust, the feasible parameter sets are $(T_{\rm d}, ~M_{\rm d}) \approx (1500 {\rm ~K}, ~ 4\times 10^{-5}{\rm ~M_{\odot}})$ for the smaller dust ($a_{\rm d} \lesssim 0.1 {\rm ~\mu m}$) and $(1600 {\rm ~K}, ~ 7\times 10^{-5}{\rm ~M_{\odot}})$ for the larger dust ($a_{\rm d} \gtrsim 1 {\rm ~\mu m}$). On the other hand, the feasible parameter sets for astronomical silicate are $(1350 {\rm ~K}, ~ 2\times 10^{-4}{\rm ~M_{\odot}})$ for $a_{\rm d} \lesssim 0.1 {\rm ~\mu m}$ and $(1350 {\rm ~K}, ~ 9\times 10^{-4}{\rm ~M_{\odot}})$ for $a_{\rm d} \gtrsim 1 {\rm ~\mu m}$. The required dust temperature for astronomical silicate is higher than its evaporation temperature; typically $\sim 1000{\rm ~K}$. Therefore, astronomical silicate is rejected as the dust species. The carbon dust is possible given its evaporation temperature of $\sim 2000 {\rm ~K}$. These results suggest that the dust is mainly composed of carbon. The estimated dust mass is roughly an order of magnitude smaller than that estimated for typical SNe IIn; this might indicate that the mass of the swept-up CSM is smaller than or at most comparable to that in SNe IIn. 

\subsubsection{The Origin of the Dust} \label{sec:5.2.2}

There are two possible origins of the NIR emission: newly-formed dust or pre-existing dust. Here we discuss the origin of the NIR thermal emission. 

First, based on an analytical model \citep{Maeda2015MNRAS,Nagao2017ApJ}, we estimate the thermal emission from pre-existing dust heated by the SN radiation, i.e., NIR echoes. Assuming radiative equilibrium at time $t$, an incoming SN flux ($L_{{\rm SN}, \nu}(t)$) and circumstellar dust temperature ($T_{\rm d}(t)$) are connected as follows:
\begin{equation}
\int_{0}^{\infty} \frac{L_{\mathrm{SN}, \nu}(t)}{4 \pi R^{2}} \kappa_{\mathrm{a}, \nu} d \nu=4 \pi \int_{0}^{\infty} \kappa_{\mathrm{a}, \nu} B_{\nu}(T_{\rm d}(t)) d \nu.
\end{equation}
We adopt the following expression for $L_{{\rm SN}, \nu}(t)$:
\begin{equation}
    L_{{\rm SN}, \nu}(t) = L_{\rm SN}(t) f_{\nu}(T_{\rm SN}(t)),
\end{equation}
where $L_{\rm SN}(t)$ is the bolometric luminosity (see Figure \ref{fig:3}) and $f_{\nu}$ is the frequency dependence of the luminosity, which is defined as follows:
\begin{equation}
    f_{\nu} = \frac{B_{\nu}(T_{\rm SN}(t))}{\int_{0}^{\infty} B_{\nu}(T_{\rm SN}(t)) \mathrm{d} \nu}
    = \frac{\pi B_{\nu}(T_{\rm SN}(t))}{\sigma T_{\rm SN}(t)^{4}},
\end{equation}
where $\sigma$ is the Stefan-Boltzmann constant and $T_{\rm SN}$ is the SN temperature. Adopting carbon dust, the dust evaporation temperature is $\sim 2000 {\rm ~K}$. Using the maximum luminosity of SN 2020uem, the dust evaporation radius ($R_{\rm eva}$) is estimated to be $\sim 0.01 {\rm ~pc}$. Then, the timescale of the echo, which is approximately equal to the light crossing time ($2R_{\rm eva}/c \sim 25 {\rm ~days}$), is too short to explain the rebrightening by the NIR echo. Therefore, the origin of the NIR emission is expected to be newly-formed dust in the cold-dense shell.

One possible concern in the newly-formed dust scenario is that it may change the wavelength-dependent optical opacity with the bluer light preferentially absorbed, which may leave a trace in the (optical) spectral evolution. In Paper II, we estimate the shock radius ($r_{\rm sh}$) of SN 2020uem at $\sim 200$ days as $\sim 1\times 10^{16} {\rm ~cm}$. Assuming a thin shell of the newly-formed carbon dust grains, the optical depth ($\tau_{\nu}(r)$) is defined as follows:
\begin{align}
    \tau_{\nu}(r) \approx \frac{\kappa_{{\rm a}, \nu} M_{\rm d}}{4\pi r_{\rm sh}^{2}}.
\end{align}
Then, the optical depth for $\rm H\alpha$ becomes around unity. In some dust-forming SNe, central wavelengths of some lines, especially evident in the optical lines, become blueshifted due to the wavelength dependence of the dust opacity \cite[e.g.,][]{Maeda2013ApJ}. However, Figure \ref{fig:6} shows that all the hydrogen lines exhibit a similar profile with a peak at the rest wavelength, which is inconsistent with the estimation of the optical depth. A similar argument applies for the $\rm H\alpha$ to $\rm H\beta$ ratio, which is expected to show a sudden increase as caused by the newly-formed dust grains; such behavior is however not seen in the spectral evolution (Section \ref{sec:6.2}). This result suggests that the dust is locally confined, e.g., torus or clump, and then the effective optical depth becomes low along the line of sight. In fact, the localized dust is also preferred in terms of the CSM geometry, which is estimated as the torus-shaped CSM in Paper II.

\section{Discussion} \label{sec:6}

In this section, we discuss differences and similarities in their observational properties between SNe IIn/Ia-CSM, including SN 2020uem, and other types of SNe (SNe Ia, IIn, and other interacting SNe).

\subsection{Light Curve Properties} \label{sec:6.1}

Although the underlying SNe in SNe Ia-CSM and SNe IIn can be completely different (i.e., thermonuclear vs. core-collapse explosions), the light curve properties (the timescale and luminosity) are very similar. This apparent contradiction might be explained as follows.

For SNe Ia-CSM, assuming a white dwarf explosion, the mass scale and energy budget are $M_{\rm{ej}}\approx 1.4 {\rm ~M_{\odot}}$ and $E_{\rm{ej}}\approx 1.0 \times 10^{51} {\rm ~erg}$, respectively. On the other hand, the acceptable regime of the mass and energy for (some) SNe IIn are $M_{\rm{ej}}\approx 10 {\rm ~M_{\odot}}$ and $E_{\rm{ej}}\approx 1.0 \times 10^{52} {\rm ~erg~s^{-1}}$ \citep[e.g.,][]{Moriya2013MNRAS_b, Dessart2015MNRAS}. Therefore, both of the mass and energy scales of SNe Ia-CSM are 10 times smaller than those of SNe IIn. Then, assuming the same swept CSM mass for both cases, the ratio of the dissipated kinetic energy (i.e., the integrated radiation energy) to the initial kinetic energy is a factor of $\sim 10$ larger for SNe Ia-CSM than SNe IIn based on the momentum conservation. In this configuration, the typical luminosity of SNe Ia-CSM is expected to be similar to those of SNe IIn. 

For SNe IIn, the swept CSM mass is expected to be at least several $M_{\rm \odot}$ \citep{Moriya2014MNRAS}. This requires that SNe Ia-CSM should have a few ${\rm M_{\odot}}$ of the CSM. To explain this CSM mass scale based on the white dwarf explosion, a likely progenitor system of SNe Ia-CSM involves a relatively low-mass star with a hydrogen envelope of a few ${\rm M_{\odot}}$, e.g., an asymptotic-giant-branch star (see also Paper II). 

\subsection{$\rm H\alpha / H\beta$ Ratio} \label{sec:6.2}

\begin{figure}
\epsscale{1.17}
\plotone{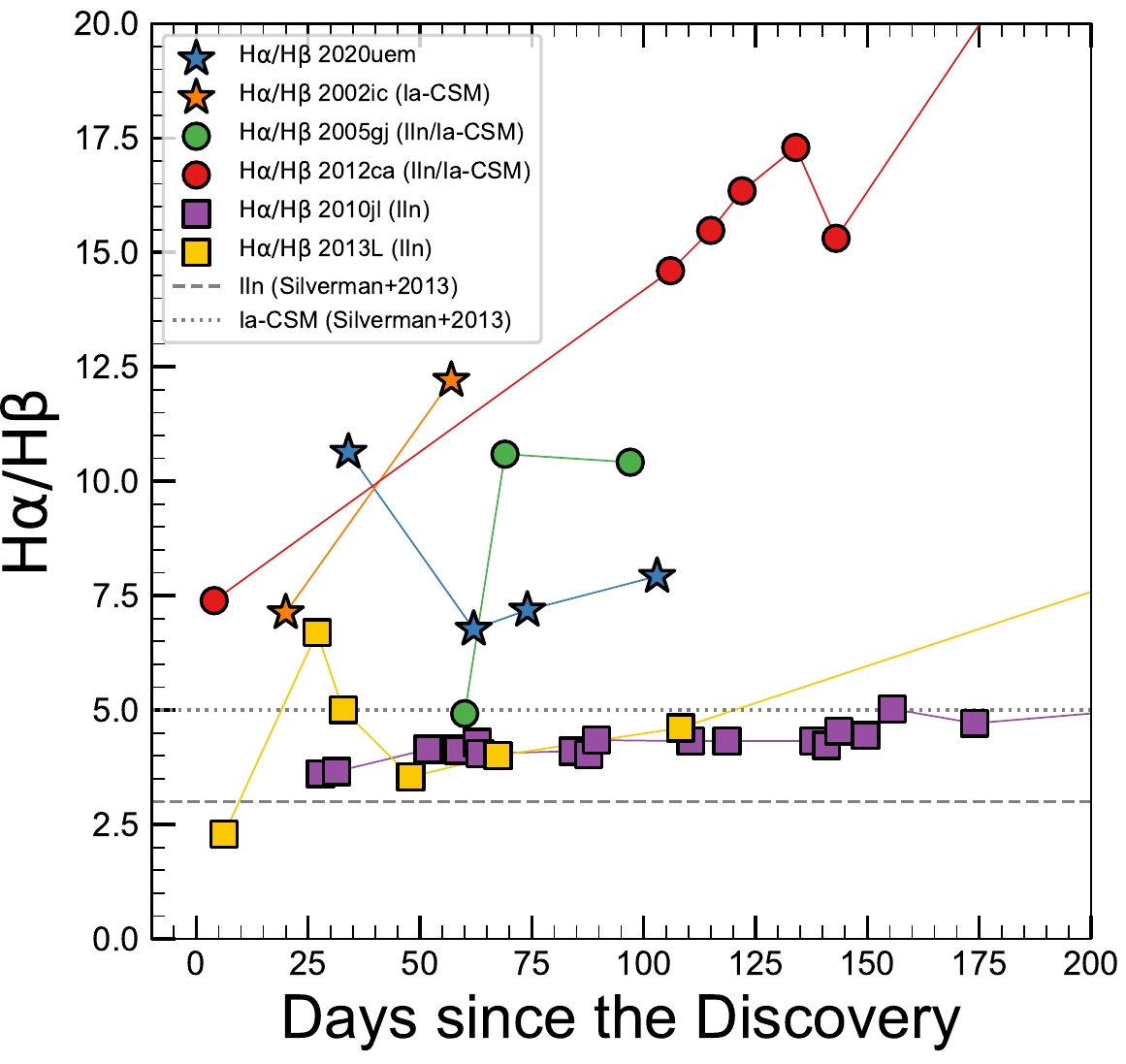}
\caption{$\rm H\alpha / H\beta$ flux ratios as a function of time for interacting SNe; SN 2013L \citep[IIn;][]{Taddia2020AA}, SN 2010jl \citep[IIn;][]{Fransson2014ApJ}, SN 2005ip \citep[IIn;][]{Stritzinger2012ApJ}, SN 2002ic \citep[Ia-CSM;][]{Hamuy2003Nature}, SN 2005gj \citep[IIn/Ia-CSM;][]{Aldering2006ApJ,Silverman2012MNRAS}, SN 2012ca \citep[IIn/Ia-CSM;][]{Inserra2014MNRAS, Fox2015MNRAS, Inserra2016MNRAS}, and SN 2020uem. The gray dashed and dotted lines show the typical values of $\rm H\alpha / H\beta$ ratios statistically investigated by \citet{Silverman2013ApJS}.}
\label{fig:12}
\end{figure}

In Figure \ref{fig:12}, we plot the time evolution of the line flux ratio of $\rm H\alpha$ to $\rm H\beta$ for SN 2020uem, together with those for other SNe IIn/Ia-CSM and IIn. The ratios for the SNe IIn/Ia-CSM, SN 2020uem and SN 2005gj, are larger than $\sim 7$, while the ratios for SNe IIn are roughly equal to $\sim 3$. Besides, we also plot the mean $\rm H\alpha / H\beta$ ratio shown by \citet{Silverman2013ApJS}, which also shows that the ratios of SNe IIn/Ia-CSM ($\sim 5$) are larger than those of SNe IIn ($\sim 3$). Therefore the high $\rm H\alpha / H\beta$ ratio seems to be a common feature in SNe IIn/Ia-CSM.

The $\rm H\alpha / H\beta$ ratio is a good tracer of the CSM density. The standard ratio is $\sim 3$ in the optically-thin limit for the Balmer series. On the other hand, the absorption of the $\rm H\beta$ line becomes non-negligible for the high-density material in which the first excited level of neutral hydrogen is highly populated. In this case, the absorbed energy which corresponds to the $\rm H\beta$ level is redistributed to the $\rm Pa\alpha$ and $\rm H\alpha$ lines. Then, the ratio can become larger than $\sim 3$. Therefore, the high $\rm H\alpha / H\beta$ ratios for SN 2020uem and other SNe IIn/Ia-CSM suggest that SNe IIn/Ia-CSM should have denser CSMs than SNe IIn\footnote{Interestingly, SN Ic-CSM 2017dio \citep{Kuncarayakti2018ApJ} shows high $\rm H\alpha / H\beta$ at the beginning, and then it goes down to around 3 later. This may support the similarity of SNe Ic-CSM (relatively compact WR star surrounded by CSM) and SNe IIn/Ia-CSM (white dwarf+CSM) in their CSM properties.}. In order to create the high-density CSM (higher than SNe IIn) for the relatively low-mass CSM (at most comparable to SNe IIn), it is required that SNe IIn/Ia-CSM should have localized CSMs, i.e., clumpy or torus-shaped CSMs.

Indeed, at least one SN Ia-CSM PTF11kx has been suggested to be associated with a disk-like CSM from a progenitor system connected to symbiotic novae \citep{Dilday2012Science}. On the other hand, \citet{Silverman2013ApJS}, who discussed the $\rm H\alpha / H\beta$ ratios in SNe Ia-CSM, suggested that the high ratios of SNe Ia-CSM are likely caused by multiple thin and dense CSM shells. However, our polarimetry data do not support such multiple shells (clumps), as will be discussed in detail in Paper II. The polarization degree and angle do not change significantly over time. This result supports an interaction scenario with a dense torus CSM rather than the collision scenario with multiple clumpy shells.

Another factor that can increase the ratio is absorption by dust. In fact, the ratio for some interacting SNe tends to be increased in late phases due to the newly-dust formation \citep[e.g.,][]{Maeda2013ApJ}. However, we consider the dust absorption is not significant for SN 2020uem since the  early-phase NIR LC does not suggest the presence of a large amount of dust (see Figure \ref{fig:2}). In addition, the amount of the newly-formed dust was likely even smaller even at $\sim 200$ days as shown in Figure \ref{fig:11}, therefore it is unlikely that the dust plays a role in the high $\rm H\alpha / H\beta$ ratio as discussed here. Indeed, late-time evolution of the $\rm H\alpha / H\beta$ ratio can provide a powerful tracer of the dust-formation process and therefore the nature of the CSM \citep[e.g.,][]{Maeda2013ApJ}. We will investigate this issue further in a  forthcoming paper with data taken in more advanced epochs than presented in the present work.

\subsection{General Spectral Features} \label{sec:6.3}

\begin{figure*}
\gridline{\fig{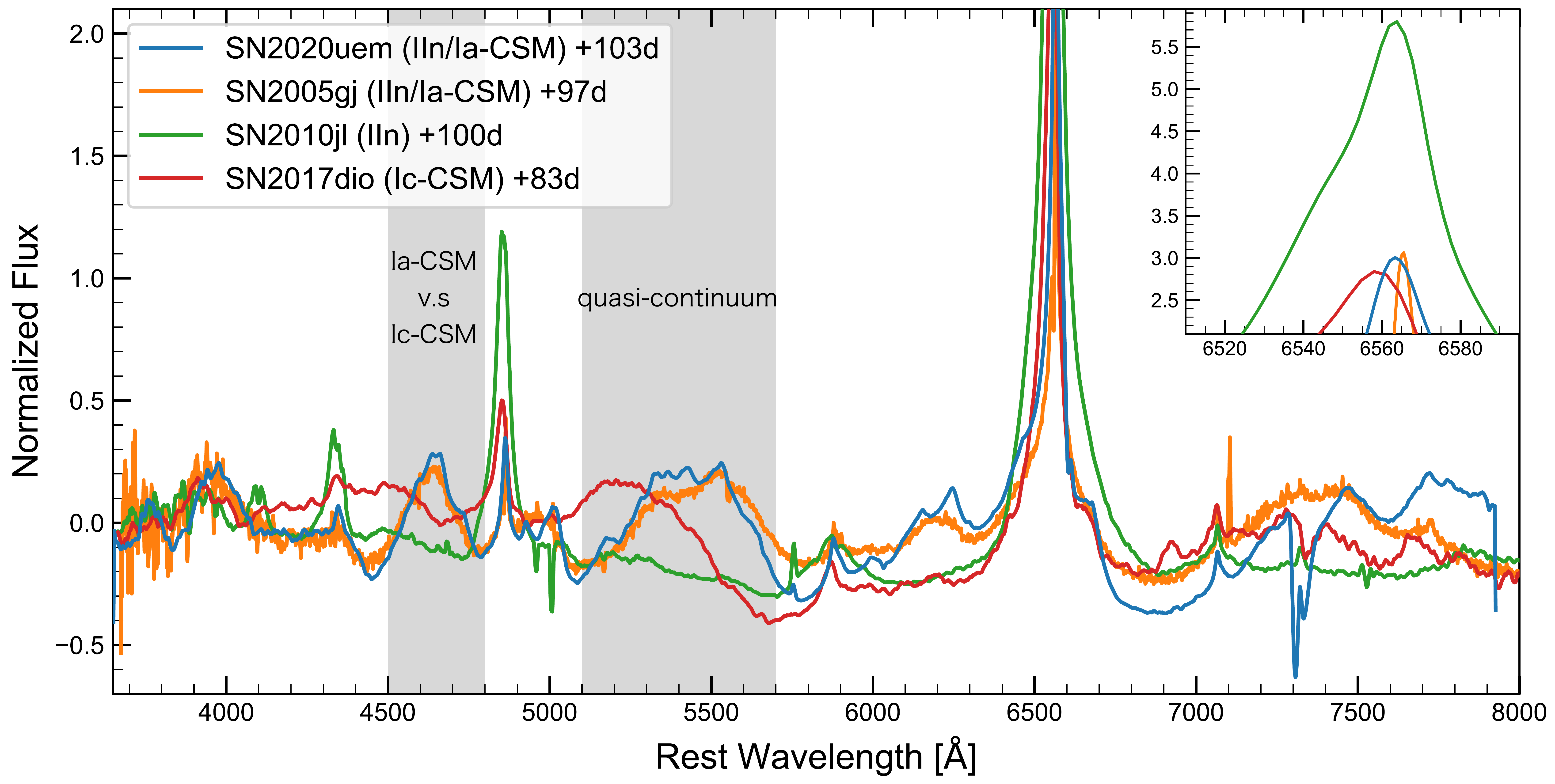}{1\textwidth}{(A) Comparison of spectra with SNe IIn/Ia-CSM and interacting SNe}
          }
\gridline{\fig{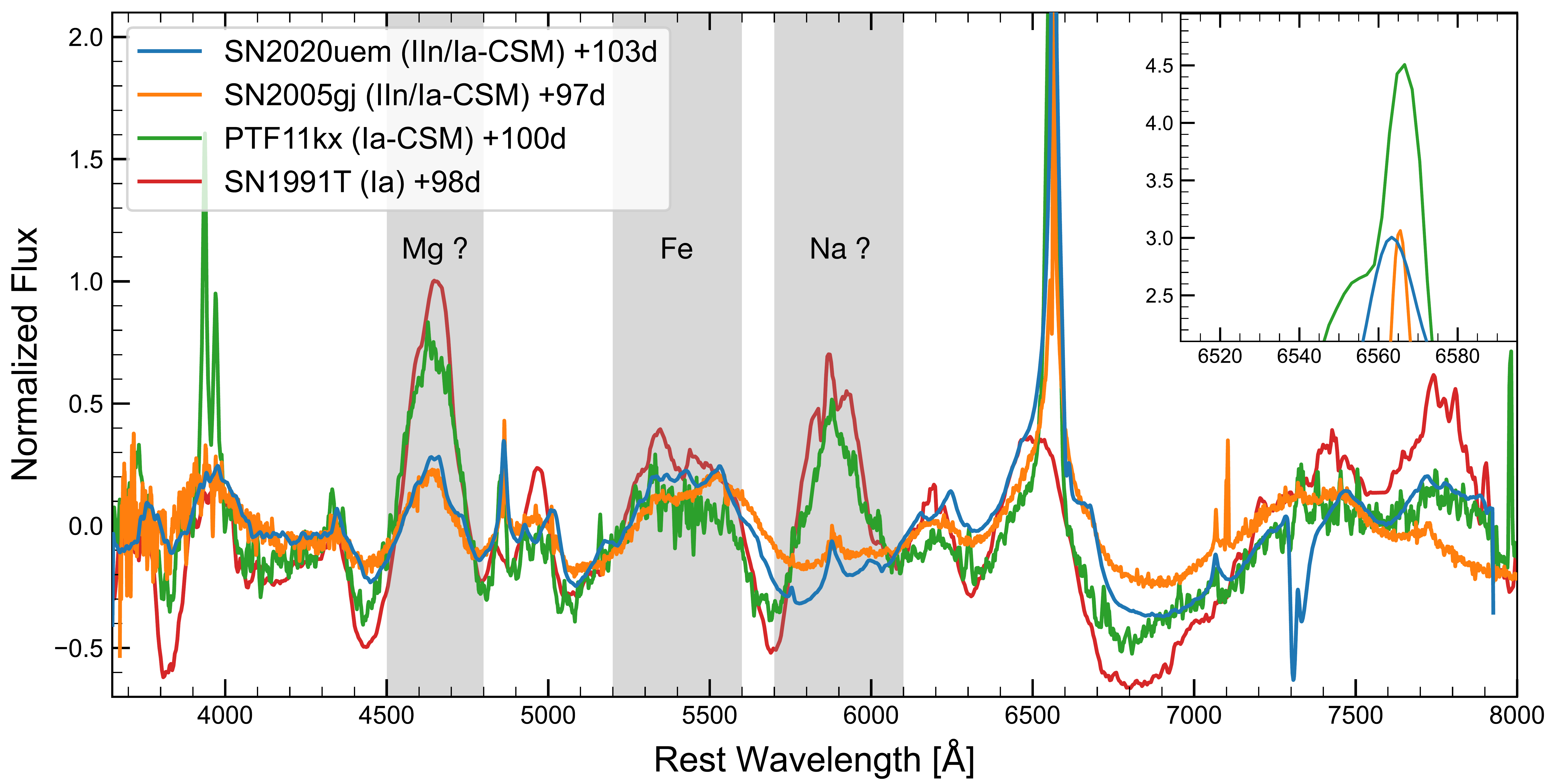}{1\textwidth}{(B) Comparison of spectra with SNe IIn/Ia-CSM and SNe Ia}
          }
\caption{The panel (A) shows a comparison of the normalized spectra of SN 2020uem and SN 2005gj \citep[IIn/Ia-CSM;][]{Silverman2012MNRAS}, SN 2010jl \citep[IIn;][]{Zhang2012AJ}, and SN 2017dio \citep[peculiar Ic;][]{Kuncarayakti2018ApJ}. The missing parts of $\rm H\alpha$ are shown in the upper right panel. The panel (B) is the same figure as the panel (A), but for PTF11kx \citep[Ia-CSM;][]{Dilday2012Science} and SN 1991T \citep[Ia;][]{Gomez1998AJ}. All spectra are normalized by their continuum using the Astropy packages. All the spectra, except for SN 2020uem, are gathered from the WISeRep. The spectral regions mentioned in the main text is highlighted with gray shade.}
\label{fig:13}
\end{figure*}

In Figure \ref{fig:13}, we compare a normalized spectrum of SN 2020uem at $\sim 100$ days after the discovery with those of other SNe at similar phases. Panel (A) shows the spectral comparison with other interacting SNe (SN 2005gj, SN 2010jl, and SN 2017dio). SN 2020uem and SN 2005gj show close similarities in their spectra. Comparing SNe IIn and IIn/Ia-CSM, the behaviors are clearly different especially at $\lesssim 6000$ {\text \AA}. The spectra of SNe IIn/Ia-CSM are dominated by a quasi-continuum possibly created by intermediate-mass elements, while the spectrum for SN 2010jl is featureless except for the Balmer series. This difference may indicate that the interaction of the SN IIn is so strong that the inner component is completely diluted by the strong continuum arising from the interaction region. Another possible factor is that SN Ia ejecta might contain richer compositions than SNe IIn, corresponding to the characteristic spectral features. The figure also shows that the $\rm H\beta$ in the SN IIn is more prominent than in the SNe IIn/Ia-CSM, reflecting the difference in the $\rm H\alpha / H\beta$ ratios between SNe IIn/Ia-CSM and SNe IIn.

Here, we also plot a normalized spectrum for a peculiar SN Ic, SN 2017dio. SN 2017dio shows H and \ion{He}{1} narrow emission lines associated with the CSM interaction, while the initial spectra are similar to SNe Ic. \citet{Kuncarayakti2018ApJ} suggested that the SN is an SN Ic exploded in a dense CSM, i.e., an SN Ic-CSM. Panel (A) shows that SN 2017dio is similar to SN 2020uem in some respects, but the spectral features around $\sim ~4650$ {\text \AA} and $\sim 5500$ {\text \AA} are different. This comparison, together with that with SNe IIn, might imply that the underlying SN behind SN 2020uem is not a core-collapse SN including SNe Ic, but a thermonuclear explosion.

Panel (B) shows the spectral comparison with non/weakly-interacting SNe Ia/Ia-CSM (SN 1991T and PTF11kx) and IIn/Ia-CSM (SN 2005gj and SN 2020uem). We focus on the features at $4500-4800$ {\text \AA}, $5200-5600$ {\text \AA}, and $5700-6100$ {\text \AA}. The feature at $5200-5600$ {\text \AA} of the SNe Ia is similar to those of the SNe IIn/Ia-CSM. On the other hand, the broad components at $4500-4800$ {\text \AA} and $5700-6100$ {\text \AA} differ between SN 1991T/PTF11kx and SN 2005gj/SN 2020uem. These features of SN 2005gj/SN 2020uem are more suppressed than those of SN 1991T/PTF11kx. These weak emission lines may indeed be a common feature of bright SNe IIn/Ia-CSM \citep[e.g., SN 2012ca and SN 2013dn;][]{Fox2015MNRAS}.

The middle-phase spectral formation ($\sim 100$ days) of SNe Ia is complicated because it is the transition phase between the photospheric phase and the nebular phase, requiring non-LTE spectral modeling. Therefore, there is still some ambiguity in the line identification. The broad feature around $5200-5600$ {\text \AA} should correspond to [\ion{Fe}{2}] and [\ion{Fe}{3}]. On the other hand, identifications of the other broad features ($4500-4800$ {\text \AA} and $5700-6100$ {\text \AA}) are unclear. Although the broad features are identified as Fe or Co lines in some SNe Ia, realistic spectral simulations do not reproduce these spectral features by iron and cobalt emissions alone \citep[e.g.,][]{Friesen2017MNRAS}. In this paper, we suggest that the broad component around $4500-4800$ {\text \AA} may be substantially contributed by \ion{Mg}{1}], while the component around $5700-6100$ {\text \AA} may be so by \ion{Na}{1}. 

We propose a speculative picture for the spectral synthesis in SNe IIn/Ia-CSM. For typical SNe Ia, according to simulations of a white dwarf explosion and nucleosynthesis \citep[e.g.,][]{Iwamoto1999ApJS}, the relatively light elements (e.g., Na, Mg, Ne, ...) are expected to be synthesized in the outer layers, while heavy elements (Fe, Co, ...) are synthesized in the inner layers. To dissipate a large fraction of the kinetic energy to explain the luminosity of SNe IIn/Ia-CSM, the reverse shock must sweep up the outer layer where Na and Mg are found. Then, it is likely that the optical, low-ionization forbidden lines from these outer elements are killed by the high temperature. In this scenario, emission lines of highly-ionized outer elements may be observed in the UV region. Note that we do not yet understand the complicated mechanisms of the spectral formation for interacting SNe. More detailed spectral synthesis calculations for interacting SNe are also important to obtain a comprehensive picture of SNe IIn/Ia-CSM.

In addition, Panel (B) clearly shows that the iron emission line of the SN Ia around $\sim 6500$ {\text \AA} overlaps with the ${\rm H\alpha}$ profile in SN 2020uem. Here, we again suggest that the asymmetric structure in the $\rm H\alpha$ profile in SN 2020uem (see also Figure \ref{fig:6}) originates from a contamination by the iron emission, and that the underlying SN of SNe IIn/Ia-CSM is likely an SN Ia.

\begin{figure}
\epsscale{1.17}
\plotone{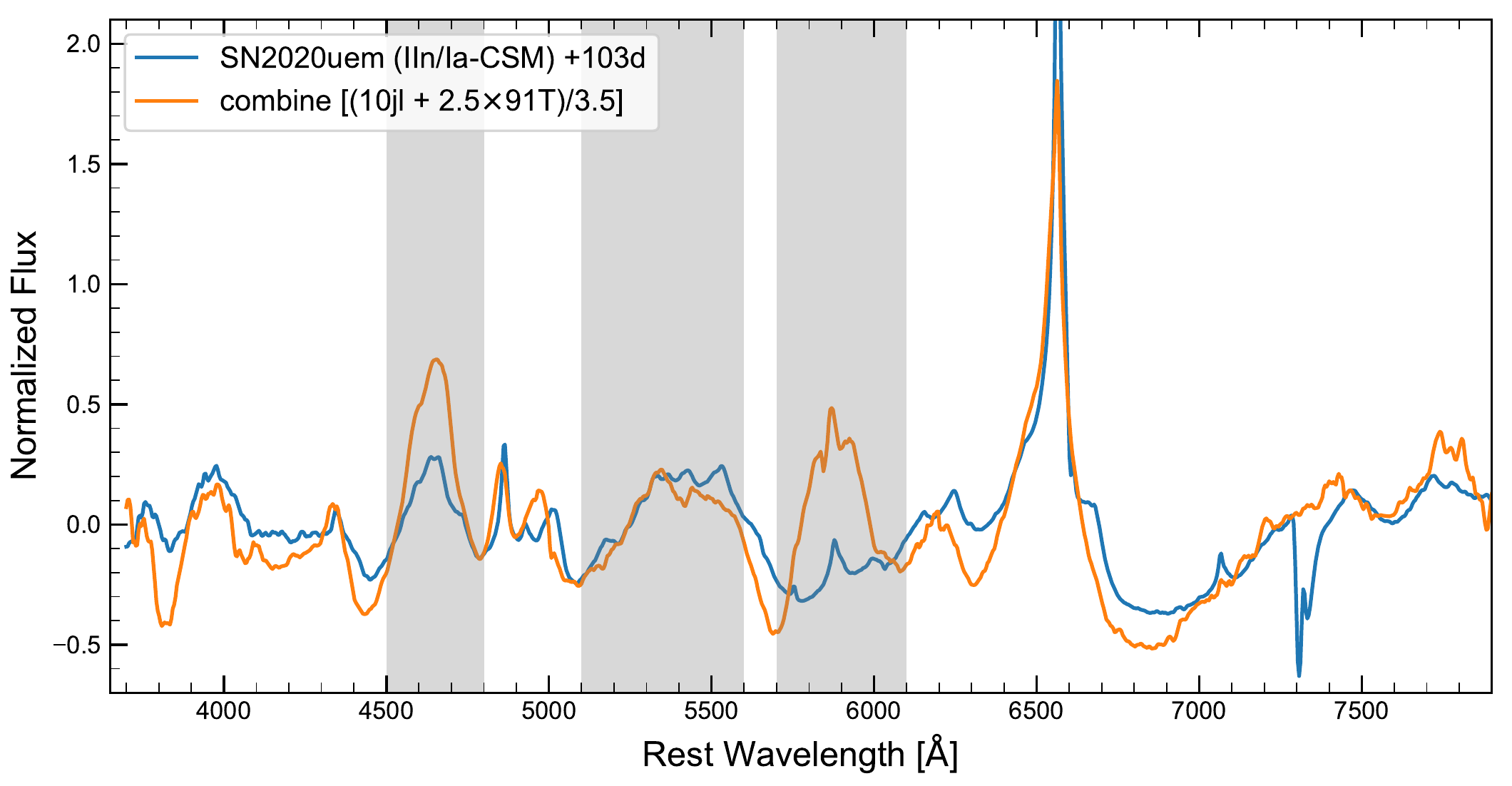}
\caption{Comparison of the normalized spectrum of SN 2020uem and a `combined' spectrum created by adding the spectra of SN 1991T and SN 2010jl. The combined spectrum is composed by $2.5\times$91T-Flux and $1\times$10jl-Flux, and additionally normalized by a factor of $3.5$. Each base spectrum in the combination is already normalized by its continuum, which is the same spectrum plotted in Figure \ref{fig:13}. The gray shaded region is the same as those in Figure \ref{fig:13}.}
\label{fig:14}
\end{figure}

To further demonstrate the similarity between SNe IIn/Ia-CSM and SNe Ia, we plot a `combined' spectrum by adding an SN IIn (2010jl) and an SN Ia (1991T) spectra in Figure \ref{fig:14}, with the relative contribution arbitrarily set following the approach by \citet{Leloudas2015AA}. The figure shows that the feature at $5200-5600$ {\text \AA} is well reproduced by the Fe lines seen in SNe Ia. Further, the other broad components ($4500-4800$ {\text \AA} and $5700-6100$ {\text \AA}) seen in SNe Ia are suppressed in the combined spectrum, being consistent with what are seen in SNe IIn/Ia-CSM. In addition, the $\rm H\alpha$ profile in the combined spectrum shows an asymmetric profile, due to the contamination of the SN-Ia component (Fe forbidden line). These results may support our scenario that the underlying SN is a thermonuclear explosion. We note however that this approach is only phenomenological, especially lacking a detailed physical mechanism of how the SN ejecta are irradiated. Also, it does not necessarily reject the core-collapse scenario if the Fe bumpy structure can be formed within the shocked CSM \citep{Chevalier2003LNP}. The detailed mechanisms of spectral formation in interacting SNe are still unclear, which is out of the scope of this paper.

\subsection{Dust Formation} \label{sec:6.4}

\begin{figure}
\epsscale{1.17}
\plotone{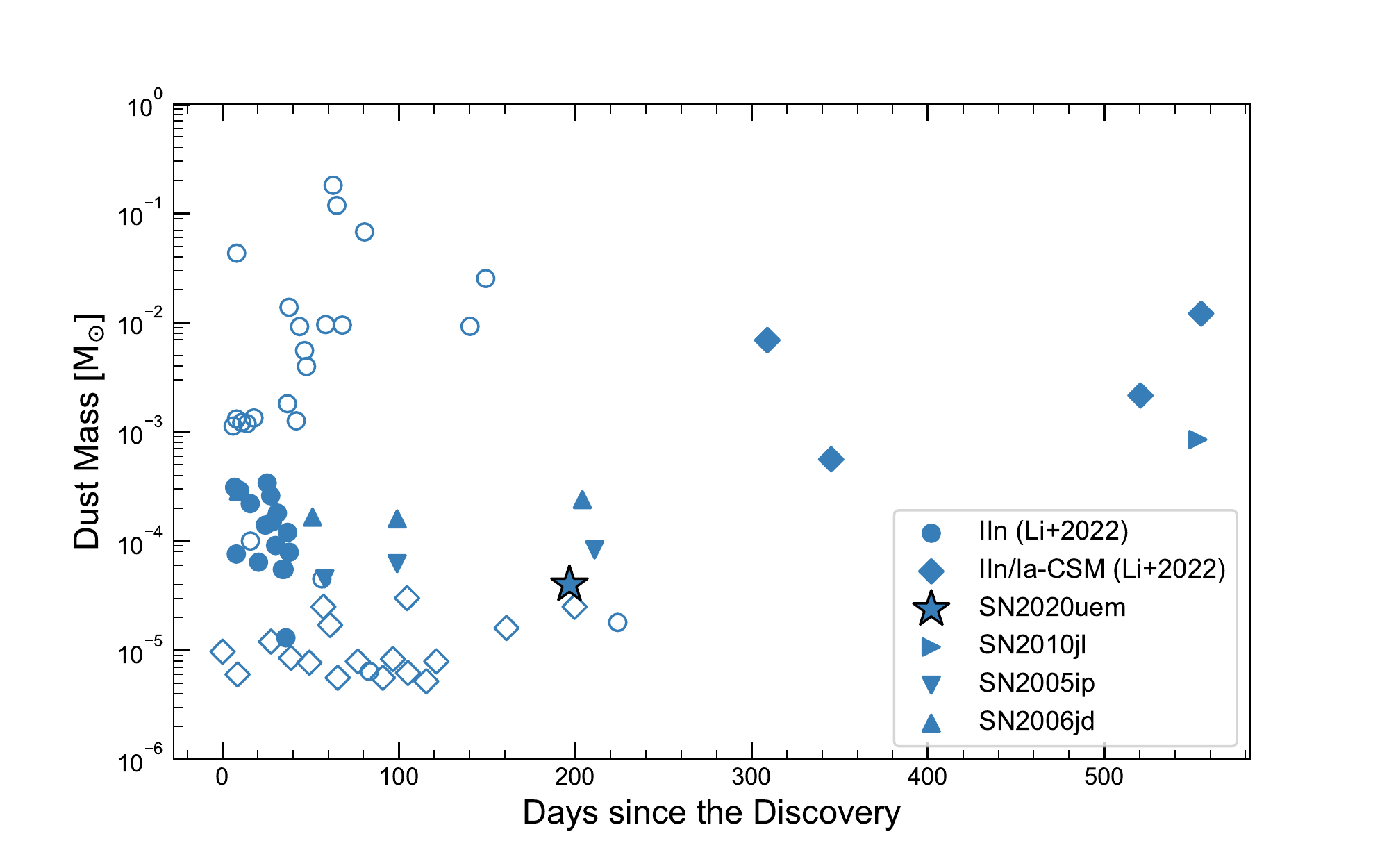}
\caption{Time evolution of the carbon dust mass. We gathered the data from \citet{Li2022ApJ}, \citet{Maeda2013ApJ} (Type IIn SN: 2010jl), and \citet{Stritzinger2012ApJ} (Type IIn SNe: 2005ip and 2006jd). The symbols are different for different spectral types of SNe; IIn (circles), IIn/Ia-CSM (diamonds), 2005ip and 2006jd (squares), and 2020uem (star). The open symbols show the upper limits of dust mass, while the filled symbols show confirmed dust mass.}
\label{fig:15}
\end{figure}

Figure \ref{fig:15} shows the comparison of the temporal evolution of the carbon dust mass for interacting SNe. In the early phase ($\lesssim 200$ days), the dust mass formed by SNe IIn is typically $\gtrsim 10^{-4}{\rm ~M_{\odot}}$ and the mass tends to be increasing with time, while that of SNe IIn/Ia-CSM is $\lesssim 10^{-4} {\rm ~M_{\odot}}$. The smaller amount of dust might reflect the difference in the progenitor system between SNe IIn/Ia-CSM and SNe IIn. Although SNe IIn/Ia-CSM and SNe IIn show similar observational features, the mass budget involved in the SN IIn/Ia-CSM system is expected to be an order of magnitude smaller than that of SNe IIn, if SNe IIn/Ia-CSM involve a white dwarf in the progenitor systems. If the gas-to-dust ratio is similar between SNe IIn/Ia-CSM and SNe IIn, the less dust mass inferred for SNe IIn/Ia-CSM might indicate a smaller total mass involved in SNe IIn/Ia-CSM, in line with the hypothesized low/intermediate-mass progenitor system.

On the other hand, the late-phase dust mass ($\gtrsim 300$ days) of SNe IIn/Ia-CSM is over $10^{-3}{\rm ~M_{\odot}}$. The mass scale is similar to other interacting SNe. We need to obtain a large sample of late-phase NIR observations of SNe IIn/Ia-CSM, in order to discuss the dust formation more quantitatively and clarify the dust formation mechanisms of interacting SNe. We expect that the data and analyses of SN 2020uem in the present work will contribute to the clarification of the dust formation in SNe IIn/Ia-CSM.

\section{Conclusions} \label{sec:7}

We have performed intensive follow-up observations of SN 2020uem, which we classify as an SN IIn/Ia-CSM, including optical/NIR photometry, spectroscopy, and polarimetry (for polarimetry, see Paper II). The multi-band photometry for SN 2020uem covers the period of $\sim 400$ days after the discovery. Our spectroscopic observations include high-dispersion and NIR spectroscopy. This is among the most massive datasets obtained for SNe IIn/Ia-CSM. The main results inferred from our dataset for SN 2020uem are as follows:

\begin{itemize}
    \item SN 2020uem shows the maximum V-band magnitude of $\lesssim -19.5$ mag. In the early phase ($\lesssim 300$ days), the light curve in each optical band evolves slowly with a decline rate of $\sim 0.75 {\rm ~mag/100~days}$. The slow light-curve evolution suggests that SN 2020uem is powered by the strong CSM interaction. On the other hand, in the late phase ($\gtrsim 300$ days), the light curve shows an accelerated decay ($\sim 1.2 {\rm ~mag/100~days}$). The accelerated decay is likely due to a change in the CSM distribution or dust formation.
    \item In the optical spectra, SN 2020uem shows prominent hydrogen emission lines and broad components by Fe-peak elements around $\sim 5500$ {\text \AA}, which are known as 'quasi-continuum' in SNe IIn/Ia-CSM. Besides, the $\rm H\alpha$ profile contains a narrow P-Cygni profile with the absorption minimum of $\sim  100 {\rm ~km~s^{-1}}$. These results also support that SN 2020uem has the slowly expanding CSM, and it is powered by the CSM interaction.
    \item The NIR spectrum of SN 2020uem exhibits narrow Paschen and Brackett emission lines and a small excess in the H- and Ks-band continuum. From the thermal excess, we estimate the dust mass and temperature as follows: $(M_{\rm d}, T_{\rm d}) \sim  (4-7 \times 10^{-5} {\rm ~M_{\odot}}, 1500-1600 {\rm ~K})$. Besides, the Ks-band light curve shows a small rebrightening around $\sim 200$ days. These properties can be interpreted by a newly-dust formation.
\end{itemize}

Based on the results, we discuss the observational features of SNe IIn/Ia-CSM and differences/similarities with SNe of other types as follows:

\begin{itemize}
    \item In order to provide the similar timescale and luminosity between SNe IIn/Ia-CSM and SNe IIn, the CSM mass of SNe IIn/Ia-CSM is likely within a few ${\rm M_{\odot}}$. This indicates that the progenitor system likely involves a low/intermediate-mass star with a hydrogen envelope of a few ${\rm M_{\odot}}$, e.g., an asymptotic-giant-branch star.
    \item The $\rm H\alpha / H\beta$ ratios for SNe IIn/Ia-CSM, including SN 2020uem, are larger than $\sim 7$, while the ratios for SNe IIn are roughly equal to $\sim 3$. This implies a higher density in the CSM around SNe IIn/Ia-CSM than SNe IIn. 
    \item Between SNe IIn/Ia-CSM and SNe Ia, there are clear differences in the spectral features at $\sim 4650$ {\text \AA} and $\sim 5900$ {\text \AA}. We suggest that the spectral features at $\sim 4650$ {\text \AA} and $\sim 5900$ {\text \AA} correspond to \ion{Mg}{1}] and \ion{Na}{1}, respectively; for SNe IIn/Ia-CSM, these elements may be suppressed due to the high ionization behind the reverse shock caused by the SN-CSM interaction.
    \item The dust mass in the early phase ($\lesssim 200$ days) of SNe IIn/Ia-CSM ($\lesssim 10^{-4}{\rm ~M_{\odot}}$) is typically more than an order of magnitude smaller than that of SNe IIn ($\gtrsim 10^{-4}{\rm ~M_{\odot}}$). This small amount of dust might support the picture that SNe IIn/Ia-CSM are triggered by white dwarf explosions, not core-collapse explosions of massive stars. Note that the late-phase dust mass ($\gtrsim 300$ days) of SNe IIn/Ia-CSM is over $10^{-3}{\rm ~M_{\odot}}$, which is similar to other interacting SNe. For more concrete discussion, we need to obtain a large sample of late-phase NIR observations for SNe IIn/Ia-CSM.
    \item The combination of the high density inferred from the $\rm H\alpha / H\beta$ ratio (higher than SNe IIn) and the CSM mass of a few $M_{\odot}$ (lower than or at most comparable to SNe IIn) inferred from the dust emission might indicate that the CSM has a larger asymmetry in SNe IIn/Ia-CSM than in SNe IIn, e.g., a more confined CSM in the former. Indeed, our polarimetric observation suggests that SN 2020uem has a torus CSM (see Paper II).
    This may hint the origin of the CSM around SNe IIn/Ia-CSM, therefore the progenitor system leading to SNe IIn/Ia-CSM. 
\end{itemize}

\begin{acknowledgments}

This research is based on observations obtained by the KASTOR (Kanata And Seimei Transient Observation Regime) team. The optical/NIR photometry and optical spectroscopy were performed by the Seimei Telescope at the Okayama observatory of Kyoto University (20B-N-CN05, 20B-E-0000, 21A-K-0001, 21A-N-CT02, 21B-N-CT09, and 21B-K-0004). The Seimei telescope is jointly operated by Kyoto University and the Astronomical Observatory of Japan (NAOJ), with assistance provided by the Optical and Near-Infrared Astronomy Inter-University Cooperation Program. We also used the data taken by the Kanata Telescope operated by the Higashi-Hiroshima Observatory, Hiroshima University. This research is also based on observations obtained by the Subaru Telescope (FOCAS: S20B-056, HDS: S21A-014, and SWIMS: S21A-019) operated by the NAOJ. The authors thank the staff at the Seimei Telescope, Kanata Telescope, and the Subaru Telescope for excellent support in the observations. The authors also thank the TriCCS developer team. The authors also thank Daichi Hiramatsu, Takashi J. Moriya, and Nozomu Tominaga for valuable discussion. The authors also thank `1st Finland-Japan bilateral meeting on extragalactic transients', which gave us a good opportunity to discuss this object. K.U. acknowledges financial support from Grant-in-Aid for the Japan Society for the Promotion of Science (JSPS) Fellows (22J22705). K.U. also acknowledges financial support from AY2022 DoGS Overseas Travel Support, Kyoto University. K.M. acknowledges support from the JSPS KAKENHI grant Nos. JP18H05223, JP20H00174, and JP20H04737. The work is partly supported by the JSPS Open Partnership Bilateral Joint Research Project between Japan and Finland (JPJSBP120229923). T.N. is funded by the Academy of Finland project 328898. T.N. acknowledges the financial support by the mobility program of the Finnish Centre for Astronomy with ESO (FINCA). Operation of SWIMS on the Subaru telescope is partly supported by Ministry of Education, Culture, Sports, Science and Technology of Japan, Grant-in-Aid for Scientific Research (20H00171) from the JSPS of Japan. H.K. was funded by the Academy of Finland projects 324504 and 328898. We have used the Weizmann interactive SN data repository to obtain the archive spectrum data for SN 1991T, SN 2002ic, SN 2005gj, PTF11kx, SN 2012ca, SN 2017dio, and SN 2010jl.

\facilities{Subaru, Seimei, Kanata}
\software{IRAF, Astropy \citep{astropy13,astropy18}}

\end{acknowledgments}

\bibliography{manuscript}{}
\bibliographystyle{aasjournal}

\movetabledown = 40mm
\begin{rotatetable}
\begin{deluxetable*}{ccccccccccc}
\tablenum{1}
\tablecaption{Log of optical photometry of SN 2020uem\label{tab:optical_photometry}}
\tablewidth{0pt}
\tabletypesize{\scriptsize}
\tablehead{
\colhead{Date} & \colhead{MJD} & \colhead{Phase} & \colhead{B} & \colhead{g} & \colhead{V} & \colhead{r} & \colhead{R} & \colhead{i} & \colhead{I} & \colhead{Telescope (Instrument)} \\
\nocolhead{} & \nocolhead{} & \colhead{(day)} & \colhead{(mag)} & \colhead{(mag)} & \colhead{(mag)} & \colhead{(mag)} & \colhead{(mag)} & \colhead{(mag)} & \colhead{(mag)} & \nocolhead{}
}
\startdata
    2020-10-27&59149.83&35.2&17.399 (0.021)&--&16.791 (0.022)&--&16.380 (0.029)&--&15.987 (0.039)&Kanata (HOWPol) \\
    2020-10-28&59150.83&36.2&17.440 (0.024)&--&16.757 (0.020)&--&16.347 (0.021)&--&15.960 (0.020)&Kanata (HOWPol) \\
    2020-11-04&59157.81&43.2&17.279 (0.026)&--&16.717 (0.022)&--&16.314 (0.021)&--&15.897 (0.021)&Kanata (HOWPol) \\
    2020-11-09&59162.75&48.1&17.358 (0.021)&--&16.724 (0.022)&--&16.288 (0.021)&--&15.862 (0.021)&Kanata (HOWPol) \\
    2020-11-14&59167.75&53.2&17.344 (0.024)&--&16.689 (0.022)&--&16.263 (0.021)&--&15.830 (0.022)&Kanata (HOWPol) \\
    2020-11-21&59174.80&60.2&17.446 (0.027)&--&16.754 (0.020)&--&16.300 (0.022)&--&15.890 (0.020)&Kanata (HOWPol) \\
    2020-11-23&59176.72&62.1&17.450 (0.023)&--&16.768 (0.020)&--&16.303 (0.020)&--&15.892 (0.021)&Kanata (HOWPol) \\
    2020-11-30&59183.77&69.2&17.518 (0.031)&--&16.879 (0.028)&--&16.395 (0.020)&--&15.991 (0.022)&Kanata (HOWPol) \\
    2020-12-06&59189.82&75.2&17.610 (0.026)&--&16.928 (0.020)&--&16.476 (0.020)&--&16.062 (0.021)&Kanata (HOWPol) \\
    2020-12-12&59195.82&81.2&17.689 (0.023)&--&17.015 (0.021)&--&16.533 (0.021)&--&16.168 (0.022)&Kanata (HOWPol) \\
    2020-12-16&59199.74&85.1&17.747 (0.020)&--&17.026 (0.026)&--&16.585 (0.021)&--&16.137 (0.027)&Kanata (HOWPol) \\
    2020-12-22&59205.79&91.2&17.822 (0.024)&--&17.161 (0.021)&--&16.699 (0.021)&--&16.311 (0.022)&Kanata (HOWPol) \\
    2020-12-30&59213.65&99.0&17.773 (0.076)&--&17.059 (0.061)&--&16.660 (0.045)&--&16.250 (0.036)&Kanata (HOWPol) \\
    2021-01-17&59231.67&117.1&17.970 (0.033)&--&17.308 (0.025)&--&16.848 (0.021)&--&16.490 (0.027)&Kanata (HOWPol) \\
    2021-01-31&59245.61&131.0&18.083 (0.043)&--&17.490 (0.025)&--&17.046 (0.027)&--&16.712 (0.021)&Kanata (HOWPol) \\
    2021-03-08&59281.61&167.0&18.265 (0.033)&--&17.763 (0.023)&--&17.315 (0.025)&--&17.034 (0.022)&Kanata (HOWPol) \\
    2021-04-01&59305.52&190.9&18.338 (0.043)&--&17.802 (0.032)&--&17.394 (0.021)&--&17.142 (0.022)&Kanata (HOWPol) \\
    2021-04-11&59315.52&200.9&18.411 (0.045)&--&17.849 (0.024)&--&17.449 (0.021)&--&17.191 (0.021)&Kanata (HOWPol) \\
    2021-04-19&59323.50&208.9&18.499 (0.074)&--&17.890 (0.025)&--&17.442 (0.036)&--&17.252 (0.043)&Kanata (HOWPol) \\
    2021-09-30&59487.82&373.2&--&19.370 (0.073)&--&19.001 (0.056)&&19.401 (0.102)&--&Seimei (TriCCS) \\
    2021-10-25&59512.78&398.2&--&19.815 (0.069)&--&19.619 (0.070)&&19.704 (0.130)&--&Seimei (TriCCS) \\
    2021-11-15&59533.80&419.2&--&19.832 (0.040)&--&19.657 (0.039)&&19.958 (0.050)&--&Seimei (TriCCS)
\enddata
\tablecomments{The Phase means days relative to the epoch of the discovery (MJD 59114.602).}
\end{deluxetable*}
\end{rotatetable}

\begin{deluxetable*}{ccccccc}
\tablenum{2}
\tablecaption{Log of NIR photometry of SN 2020uem\label{tab:NIR_photometry}}
\tablewidth{0pt}
\tabletypesize{\scriptsize}
\tablehead{
\colhead{Date} & \colhead{MJD} & \colhead{Phase} & \colhead{J} & \colhead{H} & \colhead{Ks} & \colhead{Telescope (Instrument)} \\
\nocolhead{} & \nocolhead{} & \colhead{(day)} & \colhead{(mag)} & \colhead{(mag)} & \colhead{(mag)} & \nocolhead{}
}
\startdata
    2020-10-26&59148.79&34.2&16.155 (0.030)&15.705 (0.038)&15.464 (0.078)&Kanata (HONIR) \\
    2020-10-28&59150.79&36.2&16.114 (0.031)&15.707 (0.039)&15.293 (0.070)&Kanata (HONIR) \\
    2020-10-30&59152.79&38.2&16.075 (0.029)&15.679 (0.037)&15.423 (0.066)&Kanata (HONIR) \\
    2020-11-09&59162.76&48.2&15.998 (0.035)&15.607 (0.037)&15.305 (0.065)&Kanata (HONIR) \\
    2020-11-21&59174.74&60.1&15.921 (0.024)&15.535 (0.025)&15.398 (0.043)&Kanata (HONIR) \\
    2020-11-29&59182.70&68.1&16.147 (0.032)&15.644 (0.038)&15.445 (0.072)&Kanata (HONIR) \\
    2020-11-30&59183.73&69.1&16.112 (0.032)&15.677 (0.037)&15.539 (0.066)&Kanata (HONIR) \\
    2020-12-02&59185.72&71.1&16.165 (0.023)&15.638 (0.024)&15.443 (0.039)&Kanata (HONIR) \\
    2020-12-07&59190.75&76.1&16.248 (0.029)&15.678 (0.036)&15.619 (0.062)&Kanata (HONIR) \\
    2020-12-17&59200.67&86.1&16.422 (0.031)&15.843 (0.034)&15.638 (0.056)&Kanata (HONIR) \\
    2020-12-25&59208.77&94.2&16.508 (0.041)&15.978 (0.069)&15.804 (0.125)&Kanata (HONIR) \\
    2021-01-13&59227.64&113.0&16.723 (0.047)&16.169 (0.056)&15.993 (0.125)&Kanata (HONIR) \\
    2021-01-15&59229.65&115.0& - &16.238 (0.043)&16.212 (0.130)0&Kanata (HONIR) \\
    2021-01-19&59233.63&119.0&16.775 (0.042)&16.291 (0.046)&16.270 (0.084)&Kanata (HONIR) \\
    2021-01-24&59238.62&124.0&16.965 (0.052)&16.583 (0.125)&16.383 (0.159)&Kanata (HONIR) \\
    2021-01-31&59245.63&131.0&16.940 (0.039)&16.366 (0.045)&16.137 (0.080)&Kanata (HONIR) \\
    2021-02-07&59252.67&138.1&16.935 (0.037)&16.375 (0.046)&16.404 (0.097)&Kanata (HONIR) \\
    2021-02-15&59260.66&146.1&17.088 (0.051)&16.538 (0.063)&16.541 (0.129)&Kanata (HONIR) \\
    2021-02-19&59264.55&149.9&17.083 (0.035)&16.639 (0.043)&16.672 (0.105)&Kanata (HONIR) \\
    2021-02-28&59273.58&159.0&17.023 (0.040)&16.795 (0.059)&16.567 (0.105)&Kanata (HONIR) \\
    2021-03-13&59286.55&171.9&17.093 (0.053)&16.928 (0.067)&16.428 (0.132)&Kanata (HONIR) \\
    2021-04-07&59311.26&196.7&17.439 (0.138)&-&16.490 (0.212)& Subaru (SWIMS)
\enddata
\tablecomments{The Phase means days relative to the epoch of the discovery (MJD 59114.602).}
\end{deluxetable*}

\begin{deluxetable*}{cccccc}
\tablenum{3}
\tablecaption{Log of optical/NIR spectroscopy of SN 2020uem\label{tab:spectroscopy}}
\tablewidth{0pt}
\tabletypesize{\scriptsize}
\tablehead{
\colhead{Date} & \colhead{MJD} & \colhead{Phase} & \colhead{Coverage} & \colhead{Resolution} & \colhead{Telescope (Instrument)} \\
\nocolhead{} & \nocolhead{} & \colhead{(day)} & \colhead{(\text{\AA})} & \nocolhead{} & \nocolhead{}
}
\startdata
    2020-10-14 & 59136.80 & 22.2 & 4100-8900 & 500 & Seimei (KOOLS-IFU/VPH-blue) \\
    2020-10-26 & 59148.77 & 34.2 & 4100-8900 & 500 & Seimei (KOOLS-IFU/VPH-blue) \\
    2020-11-08 & 59161.77 & 47.2 & 4100-8900 & 500 & Seimei (KOOLS-IFU/VPH-blue) \\
    2020-11-23 & 59176.78 & 62.2 & 4100-8900 & 500 & Seimei (KOOLS-IFU/VPH-blue) \\
    2020-12-05 & 59188.72 & 74.1 & 4100-8900 & 500 & Seimei (KOOLS-IFU/VPH-blue) \\
    2020-12-22 & 59205.64 & 91.0 & 4100-8900 & 500 & Seimei (KOOLS-IFU/VPH-blue) \\
    2021-01-03 & 59217.35 & 102.7 & 3650-8300 & 500 & Subaru (FOCAS/B300) \\
    2020-01-10 & 59224.64 & 110.0 & 4100-8900 & 500 & Seimei (KOOLS-IFU/VPH-blue) \\
    2021-02-06 & 59251.64 & 137.0 & 5800-8000 & 2000 & Seimei (KOOLS-IFU/VPH683) \\
    2021-03-26 & 59299.24 & 184.6 & 4400-7100 & 50000 & Subaru (HDS)\\
    2021-04-07 & 59311.29 & 196.7 & 9000-25000 & 600-1200 & Subaru (SWIMS)
\enddata
\tablecomments{The Phase means days relative to the epoch of the discovery (MJD 59114.602).}
\end{deluxetable*}

\end{document}